\documentclass[12pt]{article}
\pdfoutput=1
\usepackage{amsmath,amssymb,amsthm,bm,graphicx,float,array,multirow,rotfloat,subcaption,hyperref,cleveref,enumerate,mathdots,adjustbox,booktabs,parskip,mathtools,tikz,tikz-cd,csquotes,empheq,mathrsfs}
\usepackage[verbose,paperwidth=215.9mm,paperheight=279.4mm,tmargin=3cm,bmargin=3cm,lmargin=2.5cm,rmargin=2.5cm,footskip=1cm]{geometry}
\usepackage[para]{threeparttable}
\usepackage[all]{xy}
\usepackage[normalem]{ulem}

\numberwithin{equation}{section}
\numberwithin{figure}{section}
\allowdisplaybreaks
\makeatletter

\newif\ifpreprintoption
\@ifclasswith{revtex4-1}{preprint}{\preprintoptiontrue}{\preprintoptionfalse}
\usetikzlibrary{positioning}
\usetikzlibrary{arrows,shapes.misc}
\tikzset{gauge/.style={rounded rectangle, draw=black!100,dashed, thick, minimum size=5mm},d2/.style={rounded rectangle, draw=white!100, thick, minimum size=5mm},flavor/.style={rectangle, draw=black!100, thick, minimum size=5mm},gaugeN1/.style={rounded rectangle, draw=black!100, thick, minimum size=5mm}}

\newcommand{\veryred}[1]{{\color{red}#1_I}}
\newcommand{\veryblue}[1]{{\color{blue}#1_{II}}}

\tikzset{
node/.style={circle, thick, draw=black!100,fill=white!100,  minimum size=6mm, inner sep=0pt},
sonode/.style={circle, thick, draw=black!100,fill=red!100,  minimum size=3mm, inner sep=0pt},
spnode/.style={circle, thick, draw=black!100,fill=blue!100,  minimum size=3mm, inner sep=0pt},
fnode/.style={rectangle, thick, draw=black!100,fill=white!100,  minimum size=3mm, inner sep=0pt},
tnode/.style={rounded rectangle, outer sep=0pt, thick, minimum size=5mm}
}

\theoremstyle{plain}
\newtheorem*{thm*}{Theorem}

\theoremstyle{definition}

\newtheorem*{defn*}{Definition}

\makeatother

\theoremstyle{definition}

\begin{document}

\begin{titlepage}
\vspace*{-3cm} 
\begin{flushright}
{\tt CALT-TH-2022-012}\\
{\tt DESY-22-044}\\
{\tt UTTG 03-2022}
\end{flushright}
\vspace{1.6cm}

\begin{center}
{\LARGE\bfseries Distinguishing 6d $(1,0)$ SCFTs:\\ 
an extension to the geometric construction\\}
\vspace{1.2cm}

{\large
Jacques Distler$^{1}$, Monica Jinwoo Kang$^{2}$, and Craig Lawrie$^{3}$\\}
\vspace{.7cm}
{ $^1$ Theory Group, Department of Physics, University of Texas at Austin\\ Austin, TX 78712, U.S.A.}\par
\vspace{.2cm}
{ $^2$ Walter Burke Institute for Theoretical Physics, California Institute of Technology\\ Pasadena, CA 91125, U.S.A.}\par
\vspace{.2cm}
{ $^3$ Deutsches Elektronen-Synchrotron DESY\\ Notkestr.~85, 22607 Hamburg, Germany}\par
\vspace{.2cm}

\vspace{.3cm}

\scalebox{.8}{\tt distler@golem.ph.utexas.edu, monica@caltech.edu, craig.lawrie1729@gmail.com}\par
\vspace{1.1cm}
\textbf{Abstract}
\end{center}

\noindent 
We provide a new extension to the geometric construction of 6d $(1,0)$ SCFTs that encapsulates Higgs branch structures with identical global symmetry but different spectra. In particular, we find that there exist distinct 6d $(1,0)$ SCFTs that may appear to share their tensor branch description, flavor symmetry algebras, and central charges. For example, such subtleties arise for the very even nilpotent Higgsing of $(\mathfrak{so}_{4k}, \mathfrak{so}_{4k})$ conformal matter; we propose a method to predict at which conformal dimension the Higgs branch operators of the two theories differ via augmenting the tensor branch description with the Higgs branch chiral ring generators of the building block theories. Torus compactifications of these 6d $(1,0)$ SCFTs give rise to 4d $\mathcal{N}=2$ SCFTs of class $\mathcal{S}$ and the Higgs branch of such 4d theories are captured via the Hall--Littlewood index. We confirm that the resulting 4d theories indeed differ in their spectra in the predicted conformal dimension from their Hall--Littlewood indices. We highlight how this ambiguity in the tensor branch description arises beyond the very even nilpotent Higgsing of $(\mathfrak{so}_{4k}, \mathfrak{so}_{4k})$ conformal matter, and hence should be understood for more general classes of 6d $(1,0)$ SCFTs.

\vfill 
\end{titlepage}

\tableofcontents
\newpage

\section{Introduction}\label{sec:intro}

A generic quantum field theory is characterized by its symmetries, both global and local. Many diverse quantum field theories can be engineered from superstring theory in ten dimensions, which has no global symmetries and famously has only local symmetries required by anomaly cancellation \cite{Green:1984sg}.  However, lower dimensional theories that arise via string theory compactifications may have many kinds of global symmetries; in particular, there can be R-symmetries, if the compactification preserves supersymmetry, and flavor symmetries that commute with the (super-)Lorentz transformations. The flavor symmetry provides an important property describing the quantum field theory; understanding the flavor symmetry amounts to analyzing the flavor symmetry algebra $\mathfrak{f}$ and its global form $F$, where $F$ is ambiguous from $\mathfrak{f}$ up to the center of $\mathfrak{f}$. The spectrum of states of the theory falls into representations of $\mathfrak{f}$, and there can be subtle distinctions between the global form of the symmetry group $F$ depending on those states. The analysis of the spectrum of the theory can demonstrate that theories that appear to be identical at the level of the flavor symmetry algebra are different. Determining which states, in which representations of $\mathfrak{f}$, exist in the theory is related to the geometric and topological properties of the compactification space $Y$. However, how these states are encoded in the geometry is often challenging to determine. In this paper, we explicitly show how such states are captured in $Y$ for certain compactifications of string theory down to six dimensions and then further down to four dimensions.

In particular, the theories we look into in this paper are superconformal field theories (SCFTs). A natural question is then how does one effectively distinguish superconformal field theories. The most natural things to look at are the invariants of an SCFT. We define the ``conventional invariants'' of an SCFT to be the central charges, the flavor algebras, and the flavor central charges; these are invariants in the sense that if these quantities differ between two SCFTs, then those SCFTs are themselves different. However, these are not complete invariants: many distinct SCFTs are known for which all of these quantities are identical. A more refined invariant, but still not complete, is the Higgs branch. In another vein, the global form of the flavor symmetry group $F$ is more refined than the flavor symmetry algebra $\mathfrak{f}$, and can distinguish theories which differ only up to the center of $F$.

We will analyze six-dimensional SCFTs and take a six-dimensional perspective on analyzing four-dimensional SCFTs. In fact, studying six dimensional SCFTs has been particularly insightful and has played an important role in understanding lower-dimensional theories. The quintessential examples are the understanding of the S-duality of 4d $\mathcal{N}=4$ super Yang--Mills \cite{Witten:1995zh} and the class $\mathcal{S}$ construction \cite{Gaiotto:2009we,Gaiotto:2009hg} of 4d $\mathcal{N}=2$ SCFTs from the 6d $(2,0)$ SCFTs. The class $\mathcal{S}$ construction involves a twisted compactification of the 6d $(2,0)$ SCFT of type $\mathfrak{g}$ on a $n$-punctured genus $g$ Riemann surface $C_{g,n}$. In this paper, we write such 4d SCFTs as
\begin{equation}
    \mathcal{S}_{\mathfrak{g}}\langle C_{g,n} \rangle \{ \cdots \} \,,
\end{equation}
where $\cdots$ refers to the data describing the punctures. The puncture data has been (almost) exhaustively worked out in \cite{Wang:2015mra,Chacaltana:2010ks,Chacaltana:2011ze,Chacaltana:2013oka,Chacaltana:2014jba,Chacaltana:2015bna,Chacaltana:2016shw,Chacaltana:2017boe,Chacaltana:2018vhp,Chacaltana:2012zy}. The power of this approach is reflected in how a multitude of physical properties of the 4d SCFTs are encoded in the geometry of the punctured Riemann surface.

Another origin of 4d $\mathcal{N}=2$ SCFTs in six-dimensions is the torus compactification of a 6d $(1,0)$ SCFT. When utilizing this approach,
there is no need to perform any topological twist as the flatness of the torus guarantees that supersymmetry is preserved in the compactification. In fact, a 6d $(1,0)$ SCFT origin provides a particularly powerful perspective to understand the Higgs branch of the lower dimensional SCFTs, as a supersymmetry-preserving torus compactification does not modify the Higgs branch. This process can also be utilized in the reverse direction: if one understands aspects of the Higgs branch of a 4d $\mathcal{N}=2$ SCFT from the class $\mathcal{S}$ perspective, and there also exists a 6d $(1,0)$ on $T^2$ perspective, then one can learn about the Higgs branch of the 6d $(1,0)$ SCFT.

Four-dimensional $\mathcal{N}=2$ SCFTs which have such 6d $(1,0)$ and 6d $(2,0)$ origins have been discovered in recent years \cite{Ohmori:2015pua,DelZotto:2015rca,Ohmori:2015pia,Baume:2021qho}. The general principle is that theories of class $\mathcal{S}$ of type $\mathfrak{g}$ obtained from spheres with $N$ simple punctures and any two regular punctures associated to nilpotent orbits of $\mathfrak{g}$ have an alternative description in terms of 6d rank $N$ $(\mathfrak{g}, \mathfrak{g})$ conformal matter, Higgsed by the same nilpotent orbits of $\mathfrak{g}$, compactified on a torus. In \cite{Baume:2021qho}, it was pointed out that the 6d $(1,0)$ origin makes manifest the full flavor algebra of the 4d theory, whereas only a subalgebra is manifest in the class $\mathcal{S}$ description. This is the original example of the 6d $(1,0)$ origin being the optimal approach to the 4d Higgs branch. Unfortunately, the connection between the geometric construction of 6d $(1,0)$ SCFTs \cite{Heckman:2013pva,Heckman:2015bfa} and their Higgs branches has not been fully developed.

A vast landscape of 6d $(1,0)$ SCFTs have a geometric construction via F-theory \cite{Heckman:2013pva,Heckman:2015bfa}. This approach involves constructing the description of the theory at the generic point of its tensor branch, which is captured by a collection of curves and algebras. Furthermore, there are simple rules for building new 6d SCFTs by compositing theories associated to $\leq 3$ curves. It has generally been believed that theories with the same tensor branch description correspond to the same SCFT; in particular, such theories have the same anomaly polynomials and all attendant SCFT invariants.

However, we show that this is not the case in this paper. From the class $\mathcal{S}$ description of the $T^2$ compactification of the 6d $(1,0)$ SCFTs that we consider, we study the Hall--Littlewood index to determine the Higgs branch spectrum. By looking at the spectra via the Hall--Littlewood indices, we see that the two theories differ at somewhat large conformal dimensions, however they do have the same ``conventional invariants.'' Given that the two theories have different Higgs branches, they are necessarily distinct theories. Theories with identical conventional invariants which nevertheless differ in their Higgs branch spectrum have been studied recently in \cite{Distler:2020tub,Distler:2022nsn} for some 4d SCFTs of class $\mathcal{S}$. In terms of the 6d $(1,0)$ geometric construction, we find that there is an ambiguity in how the curves are composited together and resolving this ambiguity leads to distinct 6d SCFTs. In this way, we propose a method to recover the relevant aspects of the different Higgs branches directly from the 6d $(1,0)$ perspective, and thus provide one of the first methods to recover the higher-dimensional operators on the Higgs branch directly from the geometric construction of the 6d $(1,0)$ SCFTs.

The rest of the paper is organized as follows. In Section \ref{sec:DkDk}, we explain the construction of rank $N$ $(D_k, D_k)$ conformal matter and the Higgs branch deformations induced by pairs of nilpotent orbits of the $\mathfrak{so}_{2k} \oplus \mathfrak{so}_{2k}$ flavor symmetry from the geometric perspective of F-theory; we determine that there is a previously overlooked subtlety with the compositing by rank one $(D, D)$ conformal matter which occasionally leads to inequivalent theories with the same tensor branch description. In Section \ref{sec:examples}, we highlight these distinct theories for a variety of examples involving nilpotent orbits associated to very even D-partitions, and we determine at what conformal dimension the operator spectrum on the Higgs branch differs. The torus compactifications of these 6d $(1,0)$ SCFTs leads to 4d $\mathcal{N}=2$ SCFTs which have a dual description in class $\mathcal{S}$, and in Section \ref{sec:classs}, we observe that the two theories are also distinct from that perspective and the Hall--Littlewood index differs at the same order as predicted from the 6d $(1,0)$ description. Finally, in Section \ref{sec:conc} we conclude, discuss the significance of our results, and present some future directions.

\section{\texorpdfstring{\boldmath{$(D,D)$}}{(D,D)} conformal matter and nilpotent Higgsing}
\label{sec:DkDk}

In this paper, we provide substantial evidence that the tensor branch description of a 6d $(1,0)$ SCFT from \cite{Heckman:2013pva,Heckman:2015bfa} is insufficient, in the sense that it does not distinguish between particular 6d $(1,0)$ SCFTs that have non-isomorphic Higgs branches. Six-dimensional SCFTs are theories which contain degrees of freedom corresponding to tensionless strings \cite{Witten:1995zh,Seiberg:1996qx}, which magnetically-couple to tensor multiplets, and each of those strings acquires tension at a generic point of the tensor branch.

The geometric construction is via F-theory compactified on a non-compact elliptically-fibered Calabi--Yau threefold satisfying the negative-definite condition for the intersection pairing of compact rational curves in the base of the fibration and that the singular fibers above the intersection points of the curves are minimal.\footnote{See \cite{Heckman:2018jxk} for a recent review of the construction of 6d SCFTs from F-theory, including all necessary conditions and their derivation.}$^{,}$\footnote{If F-theory is instead compactified on a compact Calabi--Yau threefold, the resulting theory is a 6d $(1,0)$ supergravity theory. See \cite{Park:2011wv,Park:2011ji,Grimm:2012yq,DelZotto:2014fia,Esole:2015xfa,Esole:2020tby,Arras:2016evy,Esole:2017rgz,Esole:2017qeh,Monnier:2017oqd,Monnier:2018nfs,Esole:2017hlw,Esole:2018csl,Esole:2018mqb,Esole:2019asj,Esole:2014dea} for some examples of such geometric constructions.} Each compact curve gives rise to a string, with the tension proportional to the volume of the curve. The intersection pairing corresponds to the Dirac pairing on the charge lattice of the strings, and the singular fiber is associated to a gauge algebra where the gauge coupling is proportional to the inverse of the associated string tension. The SCFT limit involves taking the volume of all compact curves to zero simultaneously, which is identical to taking the tensionless limit for each string. In particular, we utilize this curve-intersection technology to build 6d $(1,0)$ SCFTs with minimal conformal matter $(G,G)$ \cite{DelZotto:2014hpa}. The geometric construction itself is modular and can be reduced to the combinatorial problem of compositing together a small collection of ``building blocks''. Specifically, we can get such a 6d SCFT from compositing together theories associated to the non-Higgsable clusters (NHCs) \cite{Morrison:1996na,Morrison:1996pp,Morrison:2012np}. Writing the negative of the self-intersection number of the curves and the algebras associated to the singular fibers, the NHCs can be written as
\begin{equation}\label{eqn:nhc}
    \begin{gathered}
      \overset{\mathfrak{su}_3}{3} \,, \quad \overset{\mathfrak{so}_8}{4} \,, \quad \overset{\mathfrak{f}_4}{5} \,, \quad \overset{\mathfrak{e}_6}{6} \,, \quad \overset{\mathfrak{e}_7}{7} \,, \quad \overset{\mathfrak{e}_7}{8} \,, \quad \overset{\mathfrak{e}_{8}}{12} \,, \quad \overset{\mathfrak{su}_{2}}{2}\overset{\mathfrak{g}_{2}}{3} \,, \quad 2\overset{\mathfrak{su}_{2}}{2}\overset{\mathfrak{g}_{2}}{3} \,, \quad \overset{\mathfrak{su}_{2}}{2}\overset{\mathfrak{so}_{7}}{3}\overset{\mathfrak{su}_{2}}{2} \,, \\
    \underbrace{2\cdots 2}_{N-1} \,, \quad 
    \underbrace{2\cdots 2}_{N-3}\overset{\displaystyle 2}{2}2 \,, \quad 
    22\overset{\displaystyle 2}{2}22 \,, \quad 
    222\overset{\displaystyle 2}{2}22 \,, \quad 
    2222\overset{\displaystyle 2}{2}22 \,.
    \end{gathered}
\end{equation}
Each NHC may be tuned, meaning that the gauge algebra can be enhanced beyond that which is written in equation \eqref{eqn:nhc}. Another key ingredient is the rank one E-string, corresponding to a $(-1)$-curve with no associated gauge algebra, and its tuned counterparts:
\begin{equation} \label{eqn:compthry}
    \overset{\mathfrak{g}}{1} \,.
\end{equation}
This theory has a flavor algebra $\mathfrak{f}$, and we can use $\overset{\mathfrak{g}}{1}$ to composite together up to two tuned  non-Higgsable clusters, for example $\overset{\mathfrak{g}_L}{n}$ and $\overset{\mathfrak{g}_R}{m}$, via gauging a $\mathfrak{g}_L \oplus \mathfrak{g}_R$ subalgebra of $\mathfrak{f}$; this would lead to
\begin{equation}
    \overset{\mathfrak{g}_L}{\vphantom{1}n}\,\,\overset{\mathfrak{g}}{1}\,\,\overset{\mathfrak{g}_R}{\vphantom{1}m} \,.
\end{equation}
As long as the resulting tensor branch configuration satisfies the negative-definiteness and minimality constraints, then one can iterate this process of composition to generate a vast landscape of 6d $(1,0)$ SCFTs. Hence each tuned E-string theory as in equation \eqref{eqn:compthry} plays a role to composite together SCFTs.

To clarify the notation, we now give an explicit example. Consider a non-compact elliptically fibered Calabi--Yau containing three compact curves in the base: $C_1$, $C_2$, and $C_3$. We take the intersection matrix to be
\begin{equation}
    C_i \cdot C_j = \begin{pmatrix}
      -1 & 1 & 0 \\
      1 & -3 & 1\\
      0 & 1 & -1 
    \end{pmatrix}_{ij} \,,
\end{equation}
where the numbers on the diagonal are the self-intersection numbers; it is straightforward to see that this matrix is negative-definite. Furthermore, take the singular fibers over each of the three curves to correspond to the gauge algebras $\mathfrak{g}_1 = \mathfrak{g}_3 = \varnothing$ and $\mathfrak{g}_2 = \mathfrak{su}_3$. Then, we can write this tensor branch configuration in a succinct form as
\begin{equation}
    1\overset{\mathfrak{su}_3}{3}1 \,.
\end{equation}
This configuration involves compositing together two copies of the rank one E-string with the $\overset{\mathfrak{su}_3}{3}$ non-Higgsable cluster. We use this concise notation throughout this work.

In this paper, we focus on the 6d $(1,0)$ SCFTs known as rank $N$ $(\mathfrak{so}_{2k}, \mathfrak{so}_{2k})$ conformal matter, and the interacting fixed points obtained by nilpotent Higgsing of the $\mathfrak{so}_{2k} \oplus \mathfrak{so}_{2k}$ flavor symmetry. Certain nilpotent Higgsings lead to theories with the same tensor branch description, however, when compactified on $T^2$ the SCFTs have an alternative description in terms of class $\mathcal{S}$, and from that perspective we see that the Higgs branches are non-isomorphic. In these cases, we propose precisely how to augment the 6d $(1,0)$ tensor branch description with additional information about the compositing theories such that we observe the distinct Higgs branches. While we focus on (nilpotent Higgsing of) $(\mathfrak{so}_{2k}, \mathfrak{so}_{2k})$ conformal matter, this is not the only occasion where an ambiguity in the compositing arises, as we discuss briefly in Section \ref{sec:conc}, and thus we expect that this additional information needs to be accounted for in the tensor branch descriptions of numerous 6d $(1,0)$ SCFTs.

The rank $N$ conformal matter theory of type $(\mathfrak{so}_{2k},
\mathfrak{so}_{2k})$ arises in M-theory as the theory living on the
worldvolume of $N$ M5-branes probing a
$\mathbb{C}^2/\Gamma_{\mathfrak{so}_{2k}}$ orbifold singularity
\cite{DelZotto:2014hpa}.  In the geometric construction of 6d $(1,0)$ SCFTs,
this theory is obtained by compositing $N-1$ copies of the tuned
non-Higgsable cluster
\begin{equation}
    \overset{\mathfrak{so}_{2k}}{4} \,,
\end{equation}
with $N$ copies of the tuned E-string:
\begin{equation}
    \overset{\mathfrak{sp}_{k-4}}{1} \,.
\end{equation}
To wit, we have the configuration
\begin{equation}\label{eqn:DDCM}
    \overbrace{\underset{[\mathfrak{so}_{2k}]}{\overset{\mathfrak{sp}_{k-4}}{1}}\ \overset{\mathfrak{\vphantom{p}so}_{2k\vphantom{-4}}}{4}\  \overset{\mathfrak{sp}_{k-4}}{1}\cdots\overset{\mathfrak{so}_{2k\vphantom{-4}}}{4}\ \underset{[\mathfrak{so}_{2k}]}{\overset{\mathfrak{sp}_{k-4}}{1}}}^{N-1 \,\,(-4)\text{-curves}} \,.
\end{equation}
We refer to the $\overset{\mathfrak{sp}_{k-4}}{1}$ as the \textit{compositing theory}, and it is an SCFT in its own right; in fact, it is the minimal $(\mathfrak{so}_{2k}, \mathfrak{so}_{2k})$ conformal matter theory. This theory has an $\mathfrak{so}_{4k}$ enhanced flavor symmetry.\footnote{The Higgs branch of this SCFT has been studied from the perspective of magnetic quivers \cite{Ferlito:2017xdq,Hanany:2018uhm,Cabrera:2019izd,Cabrera:2019dob}. Aspects of the Higgs branch of minimal
$(\mathfrak{so}_{2k}, \mathfrak{so}_{2k})$ conformal matter for some $k \geq 5$ have also been explored from a conformal bootstrap approach in \cite{Baume:2021chx}.} The
Higgs branch chiral ring has two generators: a moment map $\mu$ in the adjoint representation of the $\mathfrak{so}_{4k}$ flavor symmetry and an
additional generator $\mu^\pm$ in one of the spin representations of the $\mathfrak{so}_{4k}$. The latter transforms in the representation of the
$SU(2)$ R-symmetry with highest weight $k - 2$. A priori there can be two SCFTs, one with the Higgs branch chiral ring generated by $(\mu, \mu^+)$, and the other by $(\mu, \mu^-)$. However, it is easy to see that these are equivalent SCFTs related by the outer automorphism of $\mathfrak{so}_{4k}$. We refer to this pair of equivalent theories as
\begin{equation}\label{eqn:ambigbb}
    \overset{\mathfrak{sp}_{k-4}^+}{1} \qquad \text{ and } \qquad \overset{\mathfrak{sp}_{k-4}^-}{1} \,,
\end{equation}
respectively. This may lead us to think that the tensor branch configuration for rank $N$ $(\mathfrak{so}_{2k}, \mathfrak{so}_{2k})$ conformal matter written in equation \eqref{eqn:DDCM} is ambiguous; however, these theories are equivalent for all combinations of signs on the $(-1)$-curve. We explicitly explore this scenario and argue in Section \ref{sec:noneg} why all combinations of signs are equivalent.

The rank $N$ $(\mathfrak{so}_{2k}, \mathfrak{so}_{2k})$ conformal matter theory has an $\mathfrak{so}_{2k} \oplus \mathfrak{so}_{2k}$ flavor symmetry. Then, there exist Higgs branch renormalization group flows to new interacting fixed points, triggered by giving nilpotent vacuum expectation values to the moment map of each of the flavor symmetry factors. Let us assume that $N$ is large enough such that the nilpotent Higgsing leads to an interacting 6d SCFT. Then, we can determine the tensor branch configuration of the 6d $(1,0)$ SCFT at the end of the RG-flow from the pair of nilpotent orbits that we use to Higgs \cite{Heckman:2015ola,Heckman:2016ssk,Mekareeya:2016yal,Heckman:2018pqx,Hassler:2019eso}. Each tensor branch configuration contains compositing theories of the form
\begin{align}\label{eqn:compspq}
    \overset{\mathfrak{sp}_q}{1} \,.
\end{align}
In each case of compositions with equation \eqref{eqn:compspq}, it is necessary to determine whether there is a distinction if one composites with 
\begin{align}
    \overset{\mathfrak{sp}_q^+}{1}\quad \text{or}\quad \overset{\mathfrak{sp}_q^-}{1} \,.
\end{align} 

Nilpotent orbits of $\mathfrak{so}_{2k}$ are classified by integer partitions of $2k$, which denote the decomposition of the vector representation under the corresponding embedding of $\mathfrak{su}_2$. Since the vector representation is \emph{real}, not every partition of $2k$ is allowed: the even parts must appear with even multiplicity, yielding a \emph{D-partition}. Furthermore, each \emph{very even} D-partition -- a D-partition with only even parts -- corresponds to two distinct nilpotent orbits, which we refer to as the $\veryred{\text{red}}$ and $\veryblue{\text{blue}}$ orbits.\footnote{See e.g.~\cite{Chacaltana:2012zy} or the standard reference \cite{MR1251060} for further details on nilpotent orbits. \cite{MR1251060} uses the subscripts ``$I$" and ``$II$" to distinguish the two nilpotent orbits corresponding to a very even D-partition; \cite{Chacaltana:2012zy} uses the colors red and blue to distinguish them. Here, in a somewhat redundant notation, we will use both.} The tensor branch description after Higgsing depends only on the pair of D-partitions, and thus one concludes that the tensor branch descriptions for the Higgsings by $({\veryred{\text{red}}},{\veryred{\text{red}}})$ and $({\veryred{\text{red}}},{\veryblue{\text{blue}}})$ are the same.\footnote{There are examples in \cite{Hassler:2019eso}, where $N$ is sufficiently small, such that the $({\veryred{\text{red}}},{\veryred{\text{red}}})$ and $({\veryred{\text{red}}},{\veryblue{\text{blue}}})$ pairs of nilpotent Higgsings lead to distinct tensor branch descriptions. Such cases are exceptional.}$^,$\footnote{While the tensor branch descriptions may be identical, although it has not been found a way to see the Higgs branch operators, one may approach with the reflection of the nilpotent Higgsing in the singular geometry, corresponding to the origin of the tensor branch where all of the compact curves are shrunk to zero-volume, from T-brane dynamics \cite{Cecotti:2010bp,Anderson:2013rka,Anderson:2017rpr}.}

However, we find that while the tensor branch descriptions appear the same, the different Higgsings actually lead to theories with a different Higgs branch operator spectrum, and thus do correspond to two distinct 6d $(1,0)$ SCFTs. We see precisely for those Higgsings that the distinction between compositing with $\overset{\mathfrak{sp}_q^+}{1}$ versus $\overset{\mathfrak{sp}_q^-}{1}$ is important.

In six dimensions, an $\mathfrak{sp}_q$ gauge algebra a priori is required to be accompanied by a choice of discrete theta-angle, as $\pi_5(\mathfrak{sp}_q) = \mathbb{Z}_2$. However, if there exists $n$ hypermultiplets in the fundamental representation of $\mathfrak{sp}_q$, then the outer-automorphism of the $SO(2n)$ classical flavor symmetry rotating the hypermultiplets flips the $\theta$-angle: $0\leftrightarrow \pi$. This implies that the $\theta$-angle is rarely physically relevant \cite{Mekareeya:2017jgc,Bhardwaj:2019fzv}. The outer-automorphism also swaps the positive and negative chirality spinor representations, so it is clear that the distinction between $\mathfrak{sp}_q^\pm$ is related to the distinction between the $\theta$-angles. Exactly as for the $\mathfrak{sp}_q^\pm$, all combinations of $\theta$-angles in the quiver that are related by outer-automorphisms of the special orthogonal factors are physically equivalent. In \cite{Mekareeya:2017jgc}, the spectrum of instanton strings for an $\mathfrak{sp}_q$ gauge algebra next to an $\mathfrak{su}_{2q+8}$ gauge algebra was analyzed; there it was found that the two inequivalent embeddings of $\mathfrak{su}_{2q+8}$ inside of $\mathfrak{so}_{4q+16}$ lead to distinct string-like excitations. These two embeddings are again related to the choice of $\theta$-angle for the $\mathfrak{sp}_{q}$ gauge algebra. While we observe the distinction between the theories by studying high-dimension Higgs branch operators, it would be interesting to explore the difference between the spectra of instanton strings for the $({\veryred{\text{red}}},{\veryred{\text{red}}})$ and $({\veryred{\text{red}}},{\veryblue{\text{blue}}})$ Higgsings.

\section{Very even Higgsing in 6d via examples}\label{sec:examples}

In this section, we consider explicit examples of the 6d SCFTs that are obtained from rank $N$ $(\mathfrak{so}_{4k},\mathfrak{so}_{4k})$ conformal matter Higgsed on the left and the right by nilpotent orbits associated to very even D-partitions. In the examples that we study here, we consider Higgsing both $\mathfrak{so}_{4k}$ symmetries by nilpotent orbits associated to same very even D-partition. For the purposes of the examples in this section, we focus on the D-partitions
\begin{equation}\label{eqn:genDpart}
    [(2k-2\ell)^2, 2^{2\ell}] \,,
\end{equation}
though it is straightforward to generalize this analysis to any arbitrary pair of very even D-partitions. Each such D-partition is associated to two distinct nilpotent orbits of $\mathfrak{so}_{4k}$. As discussed, we distinguish these two orbits by coloring the D-partition red or blue and adding a subscript ``I'' or ``II''.

We give several examples to demonstrate how seemingly looking identical 6d SCFTs with identical flavor symmetry algebras are distinct and how it can be seen that they differ in their Higgs branch spectrum. The cases for the Higgsing according to equation \eqref{eqn:genDpart} where $\ell = 1$ and $\ell = 2$ are rather special, and we discuss them separately. Similarly, special care must be taken when $\mathfrak{g} = \mathfrak{so}_8$, which we study first.

\subsection{\texorpdfstring{$(\mathfrak{so}_8,\mathfrak{so}_8)$}{(so₈, so₈)} with \texorpdfstring{$2\times [2^4]$}{2×[2⁴]}}\label{sec:so8}

For our first example, we take rank $N$ $(\mathfrak{so}_8, \mathfrak{so}_8)$
conformal matter. We consider Higgsing the $\mathfrak{so}_8 \oplus
\mathfrak{so}_8$ flavor symmetry by the pairs of nilpotent orbits $(\veryred{[2^4]},
\veryred{[2^4]})$ and $(\veryred{[2^4]}, \veryblue{[2^4]})$ and contrast the two resulting theories.
The original conformal matter theory corresponds to the tensor branch
description
\begin{equation}
    1\ \overset{\mathfrak{so}_8}{4}\ 1 \underbrace{\overset{\mathfrak{so}_8}{4}\ 1\cdots\overset{\mathfrak{so}_8}{4}\ 1}_{N-3 \,\,  (-4)\text{-curves}}\overset{\mathfrak{so}_8}{4}\ 1 \,,
\end{equation}
and we assume that $N \geq 3$. According to \cite{Heckman:2016ssk}, the
tensor branch description of the SCFT obtained after the nilpotent Higgsing we
are considering is
\begin{equation}\label{eqn:firsteg}
    \underset{[\mathfrak{sp}_2]}{\overset{\mathfrak{so}_7}{3}}\ 1\underbrace{\overset{\mathfrak{so}_8}{4}\ 1\cdots\overset{\mathfrak{so}_8}{4}\ 1}_{N-3 \,\,  (-4)\text{-curves}}\underset{[\mathfrak{sp}_2]}{\overset{\mathfrak{so}_7}{3}} \,.
\end{equation}
The tensor branch description appears to be the same for both the pairs $(\veryred{[2^4]},
\veryred{[2^4]})$ and $(\veryred{[2^4]}, \veryblue{[2^4]})$ for the nilpotent Higgsing. That, however, is incorrect; the tensor branch description in equation \eqref{eqn:firsteg} is in fact ambiguous and the two possibilities correspond to distinct SCFTs. There exists
two avatars of the E-string, which have the geometric description
$\overset{\mathfrak{sp}_0^\pm}{1}$, corresponding to the Higgs branch chiral
ring possessing a generator in the positive or negative chirality spinor
representation of the $\mathfrak{so}_{16}$ flavor symmetry.\footnote{The
spinor generator for the E-string has $\Delta = 2$, and thus it combines with
the moment map operator to trigger an enhancement of the flavor symmetry
$\mathfrak{so}_{16} \rightarrow \mathfrak{e}_8$.}

We begin by studying the special case where $N = 4$, in which case the
tensor branch configuration is 
\begin{equation}\label{eqn:firstegS1}
    \underset{[\mathfrak{sp}_2]}{\overset{\mathfrak{so}_7\vphantom{0^\pm_0-4}}{3}}\ \overset{\mathfrak{sp}_0^\pm}{1}\ \overset{\mathfrak{so}_8\vphantom{0^\pm_0-4}}{4}\ \overset{\mathfrak{sp}_0^\pm}{1}\ \underset{[\mathfrak{sp}_2]}{\overset{\mathfrak{so}_7\vphantom{0^\pm_0-4}}{3}} \,.
\end{equation}
Let $\mu_1^\pm$ and $\mu_2^\pm$ denote the Higgs branch chiral ring generators
of the two E-strings, in either the positive or negative chirality spin representations. We determine the number of gauge singlets appearing in the tensor product of
these generators
\begin{equation}
    \mu_1^\pm \otimes \mu_2^\pm \,,
\end{equation}
where the tensor product is taken over the common $\mathfrak{so}_8$ gauged
subalgebra. One finds that
\begin{equation}
  \begin{aligned}
    \mu_1^+ \otimes \mu_2^+ = \mu_1^- \otimes \mu_2^- &\supset (\bm{1}, \bm{1}, \bm{1}) \,, \\
    \mu_1^+ \otimes \mu_2^- = \mu_1^- \otimes \mu_2^+ &\not\supset (\bm{1}, \bm{1}, \bm{1}) \,.
  \end{aligned}
\end{equation}
The Higgs branch generator in the spinor representation has conformal
dimension $\Delta = 2$, and thus we see that, depending on the combination of
signs in equation \eqref{eqn:firstegS1}, the SCFT may or may not have an
additional Higgs branch generator at $\Delta = 4$. Based on the comparison to
class $\mathcal{S}$, discussed in Section \ref{sec:classs}, we associate the
pairs of nilpotent orbits to tensor branch descriptions as follows:
\begin{equation}
    \begin{aligned}
      (\veryred{[2^4]},\veryred{[2^4]})\, &: \qquad \underset{[\mathfrak{sp}_2]}{\overset{\mathfrak{so}_7\vphantom{0^\pm_0-4}}{3}}\ \overset{\mathfrak{sp}_0^+}{1}\ \overset{\mathfrak{so}_8\vphantom{0^\pm_0-4}}{4}\ \overset{\mathfrak{sp}_0^+}{1}\ \underset{[\mathfrak{sp}_2]}{\overset{\mathfrak{so}_7\vphantom{0^\pm_0-4}}{3}} \,, \\
      (\veryred{[2^4]}, \veryblue{[2^4]})\, &: \qquad \underset{[\mathfrak{sp}_2]}{\overset{\mathfrak{so}_7\vphantom{0^\pm_0-4}}{3}}\ \overset{\mathfrak{sp}_0^+}{1}\ \overset{\mathfrak{so}_8\vphantom{0^\pm_0-4}}{4}\ \overset{\mathfrak{sp}_0^-}{1}\ \underset{[\mathfrak{sp}_2]}{\overset{\mathfrak{so}_7\vphantom{0^\pm_0-4}}{3}} \,.
    \end{aligned}
\end{equation}

The generalization of this analysis to $N > 4$ is now clear. In the tensor
branch configuration in equation \eqref{eqn:firsteg}, there are $N-2$
E-strings acting as compositing theories, and thus there are $N-2$ Higgs branch spinors
$\mu_i^\pm$. We must consider the gauge singlets that appear in 
\begin{equation}
    \mu_1^\pm \otimes \cdots \otimes \mu_{N-2}^\pm \,,
\end{equation}
where, again, the tensor product means that we take the tensor product
of the common $\mathfrak{so}_8$ algebras after gauging. We find two possible
options
\begin{subequations}\label{eqn:so8N}
\begin{align}
    \mu_1^+ \otimes \mu_2^+ \otimes \cdots \otimes \mu_{N-3}^+ \otimes \mu_{N-2}^+ &\supset (\bm{1,1,\cdots,1}) \,, \\
    \mu_1^+ \otimes \mu_2^+ \otimes \cdots \otimes \mu_{N-3}^+ \otimes \mu_{N-2}^- &\not\supset (\bm{1,1,\cdots,1}) \,.
\end{align}
\end{subequations}
Of course, we might expect that each of the $2^{N-2}$ combinations of signs corresponds to a different theory, however, this would represent a dramatic over-counting. Inside of the tensor branch description in equation \eqref{eqn:firsteg}, we can act by an outer-automorphism of any of the $\mathfrak{so}_8$ gauge algebras, and this has the effect of flipping the signs on the two $(-1)$-curves adjacent to that gauge algebra; as an outer-automorphism, this manifestly does not change the physical theory. We choose to use the convention that all except the left-most and right-most $(-1)$-curves have $\mu^+$; this can always be attained via a sequence of outer-automorphisms of the $\mathfrak{so}_8$ gauge algebras. In this way, we can think of the two very even nilpotent orbits as Higgsing the conformal matter theory in the following, distinct, ways:
\begin{equation}
    \begin{aligned}
        \veryred{[2^4]} &: & \quad  \overset{\mathfrak{sp}_0}{1}\ \overset{\mathfrak{so}_8}{4}\ \overset{\mathfrak{sp}_0}{1}\ \overset{\mathfrak{so}_8}{4}\ \overset{\mathfrak{sp}_0}{1}\cdots &\,\,\rightarrow\,\, \overset{\mathfrak{so}_7}{3}\ \overset{\mathfrak{sp}_0^+}{1}\ \overset{\mathfrak{so}_8}{4}\ \overset{\mathfrak{sp}_0^+}{1}\cdots \,, \\
        \veryblue{[2^4]} &: & \quad \overset{\mathfrak{sp}_0}{1}\ \overset{\mathfrak{so}_8}{4}\ \overset{\mathfrak{sp}_0}{1}\ \overset{\mathfrak{so}_8}{4}\ \overset{\mathfrak{sp}_0}{1}\cdots &\,\,\rightarrow\,\, \overset{\mathfrak{so}_7}{3}\ \overset{\mathfrak{sp}_0^-}{1}\ \overset{\mathfrak{so}_8}{4}\ \overset{\mathfrak{sp}_0^+}{1}\cdots  \,.
    \end{aligned}
\end{equation}
With this convention, it is easy to see that $(\veryred{[2^4]}, \veryred{[2^4]})$ and $(\veryblue{[2^4]}, \veryblue{[2^4]})$ give rise to the same theory after successive actions of the $\mathfrak{so}_8$ outer-automorphisms. Similarly, for all of the examples in this paper, outer-automorphisms of the $\mathfrak{so}_{2\ell}$ gauge algebras on the $(-4)$-curves can be used to show that one can always transform the combinations of signs on the compositing theories to $(+,+,\cdots,+,+)$ or $(+,+,\cdots,+,-)$. Thus, due to the two distinct combinations of signs giving rise to different numbers of gauge singlets as in equation \eqref{eqn:so8N}, we expect that the tensor branch geometry in equation \eqref{eqn:firsteg} corresponds to two distinct 6d SCFTs, which differ in their Higgs branch operator content at $\Delta = 2(N-2)$. Again, based on the matching with class $\mathcal{S}$ in Section \ref{sec:classs}, we associate the all plus SCFT to the pair of nilpotent Higgsings $(\veryred{[2^4]}, \veryred{[2^4]})$, and with one minus to $(\veryred{[2^4]}, \veryblue{[2^4]})$.

Finally, we can consider the special case where $N = 3$. The tensor branch description of the Higgsed theory is then
\begin{equation}\label{eqn:ex1}
    \overset{\mathfrak{so}_7}{3}\ 1\ \overset{\mathfrak{so}_7}{3} \,.
\end{equation}
This theory is constructed by starting with two copies of the theory
\begin{equation}
    \overset{\mathfrak{so}_7}{3} \,,
\end{equation}
and compositing together by gauging an $\mathfrak{so}_7 \oplus \mathfrak{so}_7$ subalgebra of the $\mathfrak{e}_8$ flavor symmetry of the E-string. There are two inequivalent embeddings of $\mathfrak{so}_7 \oplus \mathfrak{so}_7$ inside of $\mathfrak{e}_8$, specified by their distinct branching rules
\begin{subequations}\label{eqn:so7dec}
\begin{align}
    \mathfrak{e}_8 &\rightarrow \mathfrak{so}_7 \oplus \mathfrak{so}_7\nonumber \\
    \bm{248} &\rightarrow (\bm{21,1}) \oplus (\bm{1,21}) \oplus (\bm{7,1}) \oplus (\bm{1,7}) \oplus (\bm{8,1}) \oplus (\bm{1,8}) \label{eqn:e8toso7so7}\\ &\qquad\oplus (\bm{7,8}) \oplus (\bm{8,7}) \oplus (\bm{8,8})\, , \nonumber\\
    \mathfrak{e}_8 &\rightarrow \mathfrak{so}_7 \oplus \mathfrak{so}_7\oplus \mathfrak{u}_1\nonumber \\
    \bm{248} &\rightarrow (\bm{21,1})_0 \oplus (\bm{1,21})_0 \oplus (\bm{7,1})_2 \oplus (\bm{7,1})_{-2} \oplus (\bm{1,7})_2 \oplus (\bm{1,7})_{-2} \oplus (\bm{7,7})_0 \label{eqn:e8toso7so7u1}\\
    &\qquad \oplus (\bm{1,1})_0 \oplus (\bm{8,8})_1 \oplus
    (\bm{8,8})_{-1}\,\nonumber \,.
\end{align}
\end{subequations}
We can see that the decomposition of the moment map of the E-string contains
an $\mathfrak{so}_7 \oplus \mathfrak{so}_7$ gauge singlet in the latter branching rule given by equation \eqref{eqn:e8toso7so7u1},
whereas there is none in the former branching rule given in equation \eqref{eqn:e8toso7so7}. As this gauge singlet appears with conformal dimension $\Delta = 2$, it corresponds to a moment map operator in the gauged theory; thus, the theory with the gauge singlet has an additional $\mathfrak{u}_1$ flavor symmetry. This matches with the branching rules depicted above.

The examples in this subsection can be summarized as follows. Nilpotent Higgsing
of the $\mathfrak{so}_{8} \oplus \mathfrak{so}_8$ flavor symmetry of the rank
$N \geq 3$ $(\mathfrak{so}_{8}, \mathfrak{so}_{8})$ conformal matter theory by
the pairs of nilpotent orbits $(\veryred{[2^4]},\veryred{[2^4]})$ or $(\veryred{[2^4]}, \veryblue{[2^4]})$
leads to two distinct 6d SCFTs. These two SCFTs differ in their Higgs branch
operator spectrum starting at conformal dimension
\begin{equation}\label{eqn:eg1p5}
    \Delta = 2(N-2) \,.
\end{equation}

\subsection{\texorpdfstring{$(\mathfrak{so}_{4k},\mathfrak{so}_{4k})$}{(so₄ₖ, so₄ₖ)} with \texorpdfstring{$2\times [(2k-2)^2, 2^2]$}{2×[(2k-2)², 2²]}}\label{sec:lone}

We now consider the tensor branch configurations corresponding to Higgsing both sides of rank $N$ conformal matter of type $(\mathfrak{so}_{4k}, \mathfrak{so}_{4k})$ by one of the nilpotent orbits associated to the very even D-partition $[(2k-2)^2, 2^2]$. We assume that $k > 2$, as the $k=2$ case has been studied in Section \ref{sec:so8}. The tensor branch description of these theories is
\begin{equation}\label{eqn:ex5}
    \underset{[\mathfrak{sp}_1]}{\overset{\mathfrak{so}_{7\vphantom{-4}}}{3}}\
    \underbrace{\overset{\mathfrak{sp}_{1}}{1}\
    \overset{\mathfrak{so}_{12\vphantom{-4}}}{4}\
    \overset{\mathfrak{sp}_{3}}{1}
    \cdots \overset{\mathfrak{sp}_{2k-5}}{1}}_{k- 3\,\,  (-4)\text{-curves}}
    \overbrace{\underset{[\mathfrak{sp}_1]}{\overset{\mathfrak{so}_{4k\vphantom{-4}}}{4}}\
    \overset{\mathfrak{sp}_{2k-4}}{1}\
    \overset{\mathfrak{so}_{4k\vphantom{-4}}}{4}\
    \overset{\mathfrak{sp}_{2k-4}}{1}\cdots
    \overset{\mathfrak{so}_{4k\vphantom{-4}}}{4}\
    \overset{\mathfrak{sp}_{2k-4}}{1}\
    \underset{[\mathfrak{sp}_1]}{\overset{\mathfrak{so}_{4k\vphantom{-4}}}{4}}}^{(N+1)-2(k-1)\,\,  (-4)\text{-curves}}
    \underbrace{\overset{\mathfrak{sp}_{2k-5}}{1}\cdots
    \overset{\mathfrak{sp}_{1}}{1}}_{k-3\,\,  (-4)\text{-curves}}
    \underset{[\mathfrak{sp}_1]}{\overset{\mathfrak{so}_{7\vphantom{-4}}}{3}} \,.
\end{equation}
We require that $N \geq 2k-2$ to prevent the two nilpotent Higgsings from becoming correlated across the tensor branch.
In this quiver there are $N-2$ curves of self-intersection $(-1)$, each of which composites between the adjacent curves. Each $(-1)$-curve theory contains two Higgs branch operators: a moment map operator in the adjoint representation of the flavor symmetry, and a spinor generator in either the $S^+$ or $S^-$ representation of the flavor symmetry, as discussed around equation \eqref{eqn:ambigbb}.  We label the spinor generators of each of the $(-1)$-curve theories as $\mu_1^\pm, \mu_2^\pm, \cdots, \mu_{N-2}^\pm$. We wish to count the gauge singlets that appear in the tensor product of these spinorial generators:
\begin{equation}\label{eqn:so7mu}
    \mu_1^\pm \otimes \mu_2^\pm \otimes \cdots \otimes \mu_{N-3}^\pm \otimes \mu_{N-2}^\pm \,.
\end{equation}
We can see that the decomposition of the spinor representations of the $\mathfrak{so}_{20}$ flavor symmetries of the $\overset{\mathfrak{sp}_{1}}{1}$ compositing theories are
\begin{subequations}
\begin{align}
    \mathfrak{so}_{20} &\rightarrow \mathfrak{so}_7 \oplus \mathfrak{so}_{12} \,, \\
    S^+ &\rightarrow (\bm{1}, S^+) \oplus \cdots \,, \\
    S^- &\rightarrow (\bm{1}, S^-) \oplus \cdots \,.
\end{align}
\end{subequations}
The $\cdots$ represent terms that are not singlets under the $\mathfrak{so}_7$, and thus we can see that there are gauge singlets in the tensor product in equation \eqref{eqn:so7mu}. Depending on the combinations of signs, we find that there are two possibilities for the number of gauge singlets appearing inside of the tensor product of the spinors in equation \eqref{eqn:so7mu}:
\begin{subequations}
\begin{align}
    \mu_1^+ \otimes \mu_2^+ \otimes \cdots \otimes \mu_{N-3}^+ \otimes \mu_{N-2}^+ &\supset (\bm{1},\bm{1},\cdots,\bm{1}) \,,\\
    \mu_1^+ \otimes \mu_2^+ \otimes \cdots \otimes \mu_{N-3}^+ \otimes \mu_{N-2}^- &\not\supset (\bm{1},\bm{1},\cdots,\bm{1}) \,.
\end{align}
\end{subequations}
This indicates that the tensor branch geometry given in equation \eqref{eqn:ex5} corresponds to two 6d SCFTs that differ at conformal dimension
\begin{equation}
  2\sum_{\substack{q=1 \\ q \,\, \text{odd}}}^{2k-5} (q + 2) + (N-2k+2)(2k-2) = 2N(k-1) - 2(k-1)^2 - 2 \,,
\end{equation}
in the spectrum of Higgs branch operators. 

\subsection{\texorpdfstring{$(\mathfrak{so}_{4k},\mathfrak{so}_{4k})$}{(so₄ₖ, so₄ₖ)} with \texorpdfstring{$2\times [(2k-4)^2, 2^4]$}{2×[(2k-4)², 2⁴]}}\label{sec:ltwo}

We now turn to the case where $\ell = 2$ in the D-partition in equation \eqref{eqn:genDpart}, and furthermore we take $k \geq 4$.\footnote{The case of $k = 3$ can also be studied, but requires some modification to the exposition. We leave this as a straightforward exercise for the interested reader.} The tensor branch configuration describing the 6d SCFT(s) obtained by the nilpotent Higgsing of rank $N$ $(\mathfrak{so}_{4k}, \mathfrak{so}_{4k})$ conformal matter by nilpotent orbits associated to the very even D-partition $[(2k-4)^2, 2^4]$ is
\begin{equation}\label{eqn:ex6}
    \underset{[\mathfrak{sp}_2]}{\overset{\mathfrak{so}_{12\vphantom{-4}}}{3}}\
    \underbrace{\overset{\mathfrak{sp}_{3}}{1}\
    \overset{\mathfrak{so}_{16\vphantom{-4}}}{4}\
    \overset{\mathfrak{sp}_{5}}{1}\
    \cdots \overset{\mathfrak{sp}_{2k-5}}{1}}_{k- 4\,\, (-4)\text{-curves}}\
    \overbrace{\underset{[\mathfrak{sp}_1]}{\overset{\mathfrak{so}_{4k\vphantom{-4}}}{4}}\
    \overset{\mathfrak{sp}_{2k-4}}{1}\
    \overset{\mathfrak{so}_{4k\vphantom{-4}}}{4}\
    \overset{\mathfrak{sp}_{2k-4}}{1}\cdots
    \overset{\mathfrak{so}_{4k\vphantom{-4}}}{4}\
    \overset{\mathfrak{sp}_{2k-4}}{1}\
    \underset{[\mathfrak{sp}_1]}{\overset{\mathfrak{so}_{4k\vphantom{-4}}}{4}}}^{(N+1)-2(k-2)\,\,  (-4)\text{-curves}}\
    \underbrace{\overset{\mathfrak{sp}_{2k-5}}{1}\cdots
    \overset{\mathfrak{sp}_{3}}{1}}_{k-4\,\,  (-4)\text{-curves}}\
    \underset{[\mathfrak{sp}_2]}{\overset{\mathfrak{so}_{12\vphantom{-4}}}{3}} \,.
\end{equation}
Again, we consider the gauge singlets that appear in the tensor products of the $\mu_i^\pm$:
\begin{equation}
    \mu_1^\pm \otimes \mu_2^\pm \otimes \cdots \otimes \mu_{N-3}^\pm \otimes \mu_{N-2}^\pm \,.
\end{equation}
Of course, we can see that this would never lead to a singlet under the $\mathfrak{so}_{12}$ gauge algebras on the left and the right. However, anomaly cancellation requires that an $\mathfrak{so}_{12}$ algebra on a $(-3)$-curve includes the presence of a half-hypermultiplet in one of the spin representations of the $\mathfrak{so}_{12}$. We can consider two a priori distinct SCFTs, corresponding to $\overset{\mathfrak{so}_{12}^+}{3}$ and  $\overset{\mathfrak{so}_{12}^-}{3}$, where the sign denotes the chirality of the spinor belonging to the half-hypermultiplet. Similarly to the $\overset{\mathfrak{sp}_{q}^\pm}{1}$ theories, compositing together with different signs can lead to different SCFTs.

We refer to the scalars inside of these two half-hypermultiplets as $\mu_L^\pm$ and $\mu_R^\pm$, respectively, and then we consider gauge singlets appearing in the decomposition 
\begin{equation}\label{eqn:so12mu}
    \mu_L^\pm \otimes \mu_1^\pm \otimes \mu_2^\pm \otimes \cdots \otimes \mu_{N-3}^\pm \otimes \mu_{N-2}^\pm \otimes \mu_R^\pm \,.
\end{equation}
A priori, there is no expectation that the construction of gauge invariant operators involving hypermultiplets on the tensor branch leads to operators of the 6d SCFT at the origin. In the context of 6d SCFTs however, this is not without precedent; for example, the ``end-to-end'' operators of \cite{Razamat:2019mdt, Baume:2020ure}, are operators of 6d $(1,0)$ SCFTs obtained by taking gauge singlet combinations of hypermultiplets along the 6d quiver. Furthermore, we see that this analysis matches the alternative derivation via the class $\mathcal{S}$ construction in Section \ref{sec:classs}, and thus we have strong evidence that these operators do indeed ascend to operators of the 6d SCFT.

It is easy to see that when all of the signs in equation \eqref{eqn:so12mu} are positive one obtains a gauge singlet inside of this tensor product, and when exactly one of the signs is negative one does not obtain any gauge singlet.\footnote{Flipping any two signs leaves the number of gauge singlets invariant, as discussed in Section \ref{sec:so8}, so there are only these two distinct options.} As such, we expect that the two different combinations of signs lead to distinct 6d SCFTs, with different spectra of states on their Higgs branches. Using the known conformal dimensions of the $\mu_i^\pm$ and $\mu_{L,R}^\pm$, the difference in Higgs branch operators occurs at conformal dimension
\begin{equation}
  2+ 2\sum_{\substack{q=3 \\ q \,\, \text{odd}}}^{2k-5} (q + 2) + (N-2k+4)(2k-2) = 2kN - 2N + 8 k - 12 - 2k^2 \,.
\end{equation}

\subsection{\texorpdfstring{$(\mathfrak{so}_{4k},\mathfrak{so}_{4k})$}{(so₄ₖ, so₄ₖ)} with \texorpdfstring{$2\times [(2k-2\ell)^2, 2^{2\ell}]$}{2×[(2k-2l)², 2²ˡ]}}\label{eqn:lasteg}

Finally, we consider the cases where $\ell = 3, \cdots, k-1$, which requires that we have $k \geq 4$. The tensor branch configuration for rank $N$ $(\mathfrak{so}_{4k}, \mathfrak{so}_{4k})$ conformal matter Higgsed on the left and the right by nilpotent orbits corresponding to such a D-partition $[(2k-2\ell)^2, 2^{2\ell}]$ is
\begin{equation}\label{eqn:bigex}
{\footnotesize
    \underbrace{\overset{\mathfrak{sp}_{\ell-3}}{1}\ \overbrace{\underset{[\mathfrak{sp}_\ell]}{\overset{\mathfrak{so}_{4\ell+4}}{4}}\
    \overset{\mathfrak{sp}_{2\ell-1}}{1}\ 
    \overset{\vphantom{p}\mathfrak{so}_{4\ell+8}}{4}\ 
    \overset{\mathfrak{sp}_{2\ell+1}}{1}\ 
    \cdots \overset{\mathfrak{sp}_{2k-5}}{1}}^{k-\ell-1\,\,  (-4)\text{-curves}}
    \overbrace{\underset{[\mathfrak{sp}_1]}{\overset{\mathfrak{so}_{4k}}{4}}\ 
    \overset{\mathfrak{sp}_{2k-4}}{1}\ 
    \overset{\mathfrak{so}_{4k}}{4}\ 
    \overset{\mathfrak{sp}_{2k-4}}{1}\cdots
    \overset{\mathfrak{so}_{4k}}{4}\ 
    \overset{\mathfrak{sp}_{2k-4}}{1}\ 
    \underset{[\mathfrak{sp}_1]}{\overset{\mathfrak{so}_{4k}}{4}}}^{(N+1)-2(k-\ell)\,\,  (-4)\text{-curves}}
    \overbrace{\overset{\mathfrak{sp}_{2k-5}}{1}\cdots
    \overset{\mathfrak{sp}_{2\ell-1}}{1}
    \underset{[\mathfrak{sp}_\ell]}{\overset{\mathfrak{so}_{4\ell+4}}{4}}}^{k-\ell-1\,\,  (-4)\text{-curves}}
    \overset{\mathfrak{sp}_{\ell-3}}{1}}_{N-1 \,\,  (-4)\text{-curves}} \,.}
\end{equation}
We can see that it is necessary to have $N \geq 2(k-\ell)$ to prevent the effects of the nilpotent Higgsing on each side of the quiver from correlating with each other. The flavor algebra is
\begin{equation}
    \mathfrak{f} = \begin{cases}
      \mathfrak{sp}_{2k} \qquad &\text{ if } \ell = k - 1 \,,\,\, N = 2(k - \ell)\,, \\
      \mathfrak{sp}_\ell^{\oplus 2} \oplus \mathfrak{sp}_{2} \qquad &\text{ if } N = 2(k - \ell)\,, \\
      \mathfrak{sp}_{k}^{\oplus 2} \qquad &\text{ if } \ell = k - 1\,, \\
      \mathfrak{sp}_\ell^{\oplus 2} \oplus \mathfrak{sp}_{1}^{\oplus 2} \qquad &\text{ otherwise. }
    \end{cases}
\end{equation}
Before turning our hand to the general case, let us analyze the case with the fewest number of curves. We take
\begin{equation}
    \ell = k - 1 \qquad \text{ and } \qquad N = 2(k-\ell) = 2 \,.
\end{equation}
In this case, the Higgsing acts as follows 
\begin{equation}
  \underset{[\mathfrak{so}_{4k}]}{\overset{\mathfrak{sp}_{2k-4}}{1}}\
  \overset{\mathfrak{so}_{4k\vphantom{-4}}}{4}\
  \underset{[\mathfrak{so}_{4k}]}{\overset{\mathfrak{sp}_{2k-4}}{1}} \quad \xrightarrow{\quad([2^{2k}], [2^{2k}])\quad} \quad \overset{\mathfrak{sp}_{k-4}}{1}\
  \underset{[\mathfrak{sp}_{2k}]}{\overset{\mathfrak{so}_{4k\vphantom{-4}}}{4}}\
  \overset{\mathfrak{sp}_{k-4}}{1} \,.
\end{equation}
We expect that when the two compositing theories corresponding to the $(-1)$-curves have different chirality spinors as the generators of their chiral ring, then we will have distinct 6d SCFTs on the right-hand side. We first analyze some of the Higgs branch operator content of 
\begin{equation}\label{eqn:plusplus}
     \overset{\mathfrak{sp}_{k-4}^+}{1}\
     \underset{[\mathfrak{sp}_{2k}]}{\overset{\mathfrak{so}_{4k}\vphantom{\mathfrak{sp}_{k-4}^+}}{4}}\
     \overset{\mathfrak{sp}_{k-4}^+}{1} \,.
\end{equation}
We have four generators of the Higgs branch chiral ring before compositing: the moment maps $\mu_L$ and $\mu_R$ and the spinors $\mu_L^+$ and $\mu_R^+$. In particular, both $\mu_L^+$ and $\mu_R^+$ transform in the $S^+$ representation of their $\mathfrak{so}_{4k}$ flavor symmetries, and after gauging we find that
\begin{equation}
    \mu_L^+ \otimes \mu_R^+ \supset \bm{1}\,,
\end{equation}
where we write only the $\mathfrak{so}_{4k}$ singlet representations appearing in the decomposition. If we were to instead consider the Higgs branch of 
\begin{equation}\label{eqn:plusminus}
     \overset{\mathfrak{sp}_{k-4}^+}{1}\
     \underset{[\mathfrak{sp}_{2k}]}{\overset{\mathfrak{so}_{4k}\vphantom{\mathfrak{sp}_{k-4}^+}}{4}}\
     \overset{\mathfrak{sp}_{k-4}^-}{1} \,,
\end{equation}
then we would have $\mu_R^-$ instead of $\mu_R^+$, and the tensor product of the two different spin representations of $\mathfrak{so}_{4k}$ does not yield a singlet:
\begin{equation}
    \mu_L^+ \otimes \mu_R^- \not\supset \bm{1}\,.
\end{equation}
As the spinor generators have conformal dimension $\Delta = k - 2$ then the two theories associated to the tensor branch descriptions appearing in equations \eqref{eqn:plusplus} and \eqref{eqn:plusminus} are distinct theories, and they begin to differ in their Higgs branch spectrum at $\Delta = 2k - 4$.

It is now straightforward to consider the general tensor branch description in equation \eqref{eqn:bigex}. We can see that if all of the $N$ compositing theories have a positive chirality spinor, then there will be a gauge singlet in the $N$-fold tensor product, whereas if exactly one of the compositing theories has a negative chirality spinor then that gauge singlet is not present.\footnote{Again, flipping any pair of signs does not change the gauge singlet from what is written here.} As such, we expect these two 6d SCFTs to differ in the Higgs branch spectra at conformal dimension
\begin{equation}
  2(\ell - 1) + 2\sum_{\substack{q=2\ell-1 \\ q \,\, \text{odd}}}^{2k-5} (q + 2) + (N-2k+2\ell)(2k-2) = 2N(k-1)-2\ell -2(k-\ell )^2 \,.
\end{equation}
Due to the duality of class $\mathcal{S}$ when compactified on a torus, as discussed in Section \ref{sec:classs}, we refer to the theory with the extra gauge singlet as the Higgsing by the nilpotent orbits $(\veryred{[(2k-2\ell)^2, 2^{2\ell}]}, \veryred{[(2k-2\ell)^2, 2^{2\ell}]})$, and that without as the Higgsing by the nilpotent orbits $(\veryred{[(2k-2\ell)^2, 2^{2\ell}]}, \veryblue{[(2k-2\ell)^2, 2^{2\ell}]})$.

We now consider several special cases that will be of particular relevance in
Section \ref{sec:classs}. First, take $\ell = k - 1$, and thus the very even
D-partitions that we consider are of the form $[2^{2k}]$. We find that the
$(\veryred{[2^{2k}]}, \veryred{[2^{2k}]})$ theory has a Higgs branch operator of dimension 
\begin{equation}\label{eqn:eg1}
  (\veryred{[2^{2k}]}, \veryred{[2^{2k}]}):\ \Delta = 2(N-1)(k-1) - 2\, ,
\end{equation}
that is absent from the $(\veryred{[2^{2k}]}, \veryblue{[2^{2k}]})$ theory. Similarly, when
$\ell = k-2$ we see an operator belonging to the  Higgs branch chiral ring at
\begin{equation}\label{eqn:eg2}
  (\veryred{[4^2, 2^{2k-4}]}, \veryred{[4^2, 2^{2k-4}]}):\ \Delta = 2(N-1)(k-1) - 6 \,,
\end{equation}
in the $(\veryred{[4^2, 2^{2k-4}]}, \veryred{[4^2, 2^{2k-4}]})$ theory, that is not present in the
$(\veryred{[4^2, 2^{2k-4}]}, \veryblue{[4^2, 2^{2k-4}]})$ theory.

More generally, if we Higgs on the left with D-partition $[(2k-2\ell)^2, 2^{2\ell}]$ and on the right with D-partition $[(2k-2\ell^\prime)^2, 2^{2\ell^\prime}]$, assuming that $\ell, \ell^\prime \geq 3$ and $N \geq 2k - \ell -\ell^\prime$, we find that there is a flavor singlet Higgs branch operator in the $(\veryred{[(2k-2\ell)^2, 2^{2\ell}]}, \veryred{[(2k-2\ell^\prime)^2, 2^{2\ell^\prime}]})$ theory with conformal dimension 
\begin{equation}\label{eqn:boo}
    (\veryred{[(2k-2\ell)^2, 2^{2\ell}]}, \veryred{[(2k-2\ell^\prime)^2, 2^{2\ell^\prime}]}):\ \Delta = 2N(k-1) - (k - \ell)^2 - (k - \ell^\prime)^2 - \ell - \ell^\prime \,,
\end{equation}
which is absent in the $(\veryred{[(2k-2\ell)^2, 2^{2\ell}]}, \veryblue{[(2k-2\ell^\prime)^2, 2^{2\ell^\prime}]})$ theory. We can see that equation \eqref{eqn:boo} in fact holds more generally, when $\ell, \ell^\prime \geq 1$, by comparing to the results found in Sections \ref{sec:lone} and \ref{sec:ltwo}. In fact, by generalizing further, we can see that equation \eqref{eqn:boo} holds for $\ell, \ell^\prime \geq 0$.

\subsection{A non-example: the uniqueness of conformal matter}\label{sec:noneg}

We have now demonstrated in a variety of examples that the Higgs branch depends on whether one composites together the $(-4)$- or $(-3)$-curves with the positive or negative chirality versions of minimal $(D, D)$ conformal matter. We have observed that Higgsing by the two distinct nilpotent orbits belonging to the same very even D-partition leads to distinct 6d SCFTs. In this way, we find that the Higgs branch renormalization group flows recreate the double Hasse diagram formed by pairs of nilpotent orbits of $\mathfrak{so}_{2k}$. At this point, the reader may be wondering: why is it that only the tensor branch configurations associated to nilpotent Higgsing by very even D-partitions have two avatars? Any nilpotent Higgsing of rank $N$ $(\mathfrak{so}_{2k}, \mathfrak{so}_{2k})$ conformal matter leads to a tensor branch which contains minimal $(D, D)$ conformal matter as a compositing theory, and thus one may expect that in all cases there are distinct theories depending on whether one chooses the compositing theories to have the positive or negative chirality spinors. In this section, we demonstrate in an example that these a priori distinct theories usually give rise to the same 6d SCFT. Consider the example of rank $2$ $(\mathfrak{so}_{4k}, \mathfrak{so}_{4k})$ conformal matter, for which one can write down the following two tensor branch descriptions:
\begin{equation}\label{eqn:twoCM}
     \underset{[\mathfrak{so}_{4k}]}{\overset{\mathfrak{sp}_{2k-4}^+}{1}}\ \overset{\mathfrak{so}_{4k\vphantom{p}}}{4}\ \underset{[\mathfrak{so}_{4k}]}{\overset{\mathfrak{sp}_{2k-4}^+}{1}} \qquad \text{ and } \qquad \underset{[\mathfrak{so}_{4k}]}{\overset{\mathfrak{sp}_{2k-4}^+}{1}}\ \overset{\mathfrak{so}_{4k\vphantom{p}}}{4}\ \underset{[\mathfrak{so}_{4k}]}{\overset{\mathfrak{sp}_{2k-4}^-}{1}} \,.
\end{equation}
Once we understanding the branching rule
\begin{equation}
    \begin{aligned}
      \mathfrak{so}_{8k} &\rightarrow \mathfrak{so}_{4k} \oplus \mathfrak{so}_{4k} \\
      S^+ &\rightarrow (S^+, S^+) \oplus (S^-, S^-) \\
      S^- &\rightarrow (S^+, S^-) \oplus (S^-, S^+) \,,
    \end{aligned}
\end{equation}
it is straightforward to determine that there are the following gauge singlet states, charged under the $\mathfrak{so}_{4k} \oplus \mathfrak{so}_{4k}$ flavor symmetry, in each respective theory. In the $++$ theory we have:
\begin{equation}
    \mu_L^+ \otimes \mu_R^+ = (S^+, S^+) \oplus (S^-, S^-) \,,
\end{equation}
whereas in the $+-$ theory there is instead:
\begin{equation}
    \mu_L^+ \otimes \mu_R^- = (S^+, S^-) \oplus (S^-, S^+) \,.
\end{equation}
Thus, we see that there are the same number of gauge singlets appearing inside of $\mu_L^\pm \otimes \mu_R^\pm$, and furthermore the difference between the representations of the Higgs branch operators under the flavor symmetry can be compensated by an outer automorphism of one of the $\mathfrak{so}_{4k}$ factors. It is then clear that the two putative theories appearing in equation \eqref{eqn:twoCM} are, in fact, equivalent. For the class of 6d $(1,0)$ SCFTs obtained via nilpotent Higgsing of rank $N$ $(D, D)$ conformal matter, a general analysis, involving outer-automorphisms of the gauge and flavor algebras similar to the discussion in Section \ref{sec:so8}, reveals that there is only this subtle distinction in the Higgs branch spectrum when the tensor branch description is that associated to nilpotent Higgsing of the $\mathfrak{so}_{4k} \oplus \mathfrak{so}_{4k}$ flavor symmetry by pairs of nilpotent orbits associated to very even D-partitions.

\section{6d \texorpdfstring{\boldmath{$(1,0)$}}{(1,0)} on \texorpdfstring{\boldmath{$T^2$}}{T²} and class \texorpdfstring{\boldmath{$\mathcal{S}$}}{S}}\label{sec:classs}

At this point, the reader may be wary. We have argued for the existence of Higgs branch operators of 6d $(1,0)$ SCFTs by studying gauge invariant combinations of Higgs branch operators on the partial tensor branch. However, there is no guarantee that the operators thus-constructed actually parametrize the Higgs branch of the SCFT at the origin of the tensor branch. Indeed, the analogous construction in 4d $\mathcal{N}=2$ would fail rather badly; when one moves out on the Coulomb branch (the analogue of the tensor branch in 4d), generically the \emph{entire} Higgs branch is lifted.

Fortunately, for the classes of 6d $(1,0)$ SCFTs that we are considering, there is an alternative description of the Higgs branch at the superconformal fixed point. It is isomorphic to the Higgs branch of a certain 4d $\mathcal{N}=2$ SCFT of class $\mathcal{S}$. In the latter case, there is an independent computation of the Hilbert series of the Higgs branch, from which we can confirm our conjecture that these 6d $(1,0)$ SCFTs are distinct, despite sharing the same tensor branch description, and furthermore, vindicates our method for extracting the spectrum of Higgs branch operators from the tensor branch configuration.

To verify that our tensor branch analysis is really capturing differences in the SCFTs at the origin of the tensor branch, we use a duality to the class $\mathcal{S}$ construction \cite{Gaiotto:2009we,Gaiotto:2009hg}. It is known that rank $N$ $(\mathfrak{g}, \mathfrak{g})$ conformal matter compactified on a $T^2$ gives rise to the same 4d $\mathcal{N}=2$ SCFT as the compactification of the 6d $(2,0)$ SCFT of type $\mathfrak{g}$ on a sphere with two maximal punctures and $N$ simple punctures \cite{Ohmori:2015pua,DelZotto:2015rca,Ohmori:2015pia}. In the rank one case, this was extended beyond maximal punctures in \cite{Baume:2021qho}. We write this equivalence as
\begin{equation}\label{corresp}
    \mathcal{T}_{\mathfrak{g},N}\{O_1, O_2\}\langle T^2 \rangle = \mathcal{S}_\mathfrak{g}\langle S^2 \rangle \{O_1, O_2, O_\text{simple}^{\oplus N}\} \,.
\end{equation}
Here, $O_1$ and $O_2$ are nilpotent orbits in $\mathfrak{g}$; on the left they Higgs the $\mathfrak{g} \oplus \mathfrak{g}$ flavor symmetry in 6d, whereas in the class $\mathcal{S}$ description on the right they correspond to partial closure of the two full punctures. Due to the torus compactification, the Higgs branch of this 4d $\mathcal{N}=2$ SCFT is identical to the Higgs branch of the original 6d $(1,0)$ theory.

The Hall--Littlewood limit of the superconformal index \cite{Gadde:2011uv,Kinney:2005ej,Gadde:2009kb,Gadde:2011ik,Lemos:2012ph} can be obtained from the class $\mathcal{S}$ description. It is a formal power series of the form
\begin{equation}
I_{\text{HL}}(\tau)= \text{Tr}_{\mathcal{H}_{\text{HL}}}\tau^{2(\Delta-R)} {(-1)}^F \,,
\end{equation}
where $\mathcal{H}_{\text{HL}}$ is the subspace of local operators satisfying $\Delta-2R-r=j_1=0$; here $\Delta$ is the conformal dimension, $R$ is the charge under the $SU(2)$ R-symmetry, and $r$ the charge under the $U(1)$ R-symmetry. The index counts (with sign) operators in short multiplets of the superconformal symmetry, $\hat{B}_R$ and $\mathcal{D}_{R(0,j_2)}$ (in the notation of Dolan and Osborn \cite{Dolan:2002zh}). It is generally believed \cite{Beem:2014rza,Gadde:2011uv,Beem:2013sza} that there are no $\mathcal{D}_{R(0,j_2)}$ multiplets in genus-zero theories of class $\mathcal{S}$.\footnote{Exceptions to this conjecture seem to occur for class $\mathcal{S}$ on spheres with at least four twisted-punctures \cite{HLneqHS}. As we do not consider such configurations in this paper, these exceptions are not relevant.} In which case, the Hall--Littlewood index coincides with the Hilbert series of the Higgs branch of the class $\mathcal{S}$ theory, with each $\hat{B}_R$ operator contributing $\tau^{2R}$ to the index. The refined  version of the index, $I_{HL}(\textbf{a};\tau)$ is defined similarly but with the coefficient of $\tau^{2R}$ being the character $\chi(a)_R$ of the flavor symmetry representation under which the $\hat{B}_R$ operators transform, rather than merely the dimension. 

For an $(N+2)$-punctured sphere, the Hall--Littlewood index takes the form \cite{Gadde:2011uv,Lemos:2012ph}
\begin{equation}\label{IHL}
	I_{HL}(\textbf{a};\tau) = \sum_\Lambda \frac{\prod_{i=1}^{N+2} \mathcal{K}_{HL}(\textbf{a}_i)P_\Lambda(\textbf{a}_i)}{\left(\mathcal{K}_{HL}(\{\tau\})P_\Lambda(\{\tau\})\right)^N} \,,
\end{equation}
where we describe each term contributing to this expression below.
\begin{enumerate}
	\item The sum is over finite dimensional irreducible representations $\Lambda$ of $\mathfrak{g}$. Here, we are interested in $\mathfrak{g}=\mathfrak{so}_{4k}$ and we denote $\Lambda$ by its Dynkin labels 
	\begin{equation}
	    \Lambda=(n_1,n_2,\dots,n_{n-2};n_{S^+},n_{S^-}) \,,
	 \end{equation}
	 where the last two Dynkin labels are those associated to the two irreducible spinor representations. The outer-automorphism of $\mathfrak{so}_{4k}$ acts on the representation ring by exchanging $n_{S^+}\leftrightarrow n_{S^-}$.
	
	\item Flavor fugacities $\textbf{a}_i$ associated to the $i^{\text{th}}$ puncture are determined by decomposition of the fundamental representation of $\mathfrak{g}$ as a representation of $\rho_i(\mathfrak{su}_{2})\times \mathfrak{f}_i$. Here, $\rho_i : \mathfrak{su}_2 \rightarrow \mathfrak{so}_{4k}$ describes the embedding associated to the nilpotent orbit of the $i^\text{th}$ puncture; the $\mathfrak{f}_i$ is the remaining flavor symmetry algebra.
	Furthermore, $\{\tau\}$ is the fugacity for the trivial puncture (i.e. the \emph{regular embedding} of $\mathfrak{su}_2\hookrightarrow \mathfrak{so}_{4k}$).
	
	\item The $\mathcal{K}$-factor associated to the $i^{\text{th}}$ puncture is determined by the restriction of the adjoint representation $\text{ad}_\mathfrak{g}$ of $\mathfrak{g}$ to $\rho_i(\mathfrak{su}_{2})\times \mathfrak{f}_i$ as
	\begin{align}
	\text{ad}_{\mathfrak{g}}= \bigoplus_j V_j \otimes R_{j,i} \,,
	\end{align}
	where $V_j$ is the $(2j+1)$-dimensional irreducible representation of $\mathfrak{su}_{2}$ and $R_{j,i}$ is the corresponding representation of $\mathfrak{f}_i$, possibly reducible. Upon this decomposition, a $\mathcal{K}$-factor for the $i^\text{th}$ puncture is
	\begin{equation}
		\mathcal{K}_{HL}(\textbf{a}_i)= (1-\tau^2)^{\frac{\text{rank}(\mathfrak{g})}{2}}\text{PE}\left[\sum_j \tau^{2(j+1)} \chi^{\mathfrak{f}_i}_{R_{j,i}}(\textbf{a}_i)\right] \,,
	\end{equation}
	where PE$[\cdots]$ denotes the plethystic exponential and $\chi^{\mathfrak{f}_i}_{R_{j,i}}(\textbf{a}_i)$ is the character of the flavor algebra $\mathfrak{f}_i$ in the relevant representation.
	
	\item The $P_\Lambda$ are Hall--Littlewood polynomials for the representation $\Lambda$,  which are given by
	\begin{align}
		P_\Lambda (\textbf{a}_i) &=\frac{1}{W_{\Lambda}(\tau)} \sum_{w\in W} e^{w(\Lambda )} \prod_{\alpha \in \Phi_+}\frac{1- \tau^2 e^{- w(\alpha)}}{1-e^{-w(\alpha)}} \,,\\
		W_\Lambda(\tau)&= \sqrt{\sum_{w \in \text{Stab}_W(\Lambda)}\tau^{2l(w)}} \,,
\end{align}
where $\Phi_+$ are the positive roots of $\mathfrak{g}$, $W$ is the Weyl group of $\mathfrak{g}$, and flavor fugacities $\{\textbf{a}_i\}$ can be assigned once we choose a basis for the weight lattice for $\mathfrak{g}$. 
\item The unrefined index is recovered in the limit of setting the fugacities $a_i\to1$. 
\end{enumerate}

The simple puncture in the class $\mathcal{S}$ theory of type $\mathfrak{so}_{4k}$ corresponds to the D-partition $[4k-3,3]$. For the punctures $O_1$, $O_2$, we will take two very even D-partitions. It is a fundamental fact of the representation theory of $\mathfrak{so}_{4k}$ that a very-even D-partition, $O$, gives rise to two nilpotent orbits, $\veryred{O}$ and $\veryblue{O}$. 
The two orbits are exchanged by the outer-automorphism of $\mathfrak{so}_{4k}$, which exchanges $S^+\leftrightarrow S^-$.

The spinor representations $S^+$ and $S^-$ decompose differently under the corresponding embeddings of $\mathfrak{su}_2\oplus \mathfrak{f}$.
\begin{itemize}
\item Under the embedding corresponding to $\veryred{O}$:
\begin{itemize}
\item For $k$ even, $S^+$ decomposes as half-integer-spin representations of $\mathfrak{su}_2$ tensored with pseudoreal representations of $\mathfrak{f}$, while $S^-$ decomposes as a direct sum of integer-spin representations of $\mathfrak{su}_2$ tensored with real representations of $\mathfrak{f}$.
\item For $k$ odd, $S^+$ decomposes as half-integer-spin representations of $\mathfrak{su}_{2}$ tensored with real representations of $\mathfrak{f}$, while $S^-$ decomposes as a direct sum of integer-spin representations of $\mathfrak{su}_{2}$ tensored with pseudoreal representations of $\mathfrak{f}$.
\end{itemize}
\item Under the embedding corresponding to $\veryblue{O}$, the decompositions of $S^+$ and $S^-$ are reversed. 
\end{itemize}
By contrast, representations of the form $\Lambda=(n_1,n_2,\dots,n_{2k-2};0,0)$ decompose identically under the embeddings corresponding to $\veryred{O}$ and $\veryblue{O}$.

The outer-automorphism of $\mathfrak{so}_{4k}$ is clearly a symmetry of the conformal field theory. In the index in equation \eqref{IHL}, it exchanges $\Lambda^+$ with $\Lambda^-$, that are given by
\begin{align}
    \Lambda^+=(n_1,n_2,\dots,n_{2k-2};n_{S^+},n_{S^-}),\quad \Lambda^-=(n_1,n_2,\dots,n_{2k-2};n_{S^-},n_{S^+});
\end{align}
i.e.~the representations with the final two Dynkin labels interchanged. This obviously leaves the sum unchanged. In particular, this means that the theory with two very even punctures $\veryred{O},\; \veryred{O'}$ is isomorphic to the theory with $\veryblue{O},\; \veryblue{O'}$. However, they are not necessarily (and, in general, are not) isomorphic to the theory with $\veryred{O},\; \veryblue{O'}$. At the level of the index in equation \eqref{IHL}, however, the difference is invisible up to the order in the $\tau$-expansion at which the representations $\Lambda=(0,0,\dots,0;1,0)$ and/or $(0,0,\dots,0;0,1)$ first contribute to the index.

There is a simple formula \cite{Chacaltana:2018vhp} for the order at which each representation $\Lambda$ first contributes to the index written in equation \eqref{IHL}. Let 
\begin{equation}
w(O)=(w_1,w_2,\cdots,w_{2k-2};w_{S^+},w_{S^-})
\end{equation}
be the weighted Dynkin diagram corresponding to the nilpotent orbit $O$.\footnote{See page 83 of \cite{MR1251060} for the determination of the weighted Dynkin diagram from the D-partition.} The entries of $w(O)$ are either $0,1$ or $2$. For the simple puncture, we have 
\begin{equation}
    w([4k-3,3])=(2,2,\dots,2,0;2,2) \,,
\end{equation} 
and the weighted Dynkin diagram corresponding to the trivial puncture ($O=O_{\text{regular}}$) is 
\begin{equation}
    w_0=(2,2,\dots,2;2,2) \,.
\end{equation}
Then, the leading contribution to the index from the representation $\Lambda$ is the contribution from the characters $\chi_{R_\Lambda}(\textbf{a})\tau^{n_\Lambda}$, where 
\begin{equation}\label{nLambdadef}
    n_\Lambda=\Lambda\cdot C^{-1}\cdot\Bigl(N w_0 - \sum_{i=1}^{N+2} w(O_i)\Bigr) \,,
\end{equation}
and $C$ is the Cartan matrix. The character $\chi_{R_\Lambda}$ is obtained as follows. First, for each puncture $O_i$, we can decompose the representation $\Lambda$ of $\mathfrak{so}_{4k}$ under the corresponding embedding $\rho_i(\mathfrak{su}_{2})\times \mathfrak{f}_i$ as
\begin{equation}\label{Vdecomp}
    \Lambda=\bigoplus_j V_j\otimes R_{\Lambda\;j,i} \,.
\end{equation}
Next, we let $j_i$ be the largest value of $j$ that occurs in the decomposition in equation \eqref{Vdecomp} at the $i^{\text{th}}$ puncture. Then, 
\begin{equation}\label{Rlambdadef}
  R_\Lambda=\bigotimes_{i=1}^{N+2} R_{\Lambda\;j_i,i} \,.
\end{equation}
We briefly highlight this with an example. Consider $\mathfrak{g}=\mathfrak{so}_8$ with three punctures: 
$\veryred{[2^4]}$, $\veryred{[2^4]}$, and $[5,3]$. For $\Lambda = \bm{8_v}$ we have the following decompositions:
\begin{equation}
  \begin{aligned}
    &\veryred{[2^4]} \,: \begin{cases} \begin{aligned}
        \mathfrak{so}_8 &\rightarrow \mathfrak{su}_2 \oplus \mathfrak{sp}_2 \,, \\
        \bm{8_v} &\rightarrow (\bm{2}, \bm{4}) \,,
    \end{aligned}
    \end{cases} \\
    &[5,3] \,: \begin{cases} \begin{aligned}
        \mathfrak{so}_8 &\rightarrow \mathfrak{su}_2 \,, \\
        \bm{8_v} &\rightarrow \bm{5} \oplus \bm{3} \,.
    \end{aligned}
    \end{cases}
    \end{aligned}
\end{equation}
As such, we can see that $R_{\Lambda = \bm{8_v}} = (\bm{4}, \bm{4})$ under the $\mathfrak{sp}_2 \oplus \mathfrak{sp}_2$ flavor symmetry from the two $\veryred{[2^4]}$ punctures. 

\begin{figure}[H]
\begin{subfigure}{0.5\textwidth}
\begin{center}
    \begin{tikzpicture}
        \shade[inner color=white, outer color=blue!40, opacity = 0.4] (0,0) circle (1.4cm);
        \draw (0,0) circle (1.4cm);
        \draw (-1.4,0) arc (180:360:1.4 and 0.6);
        \draw[dashed] (1.4,0) arc (0:180:1.4 and 0.6);
        \draw[thick,fill=white] (-.6,.75) circle (3pt);
        \node[font=\scriptsize] (P1) at (-.1,.8) {$\veryred{[2^{2k}]}$};
        \draw[thick,fill=white] (0,.1) circle (3pt);
        \draw[thick,fill=black] (.2,.1) circle (.5pt);
        \draw[thick,fill=black] (.3,.1) circle (.5pt);
        \draw[thick,fill=black] (.4,.1) circle (.5pt);
        \draw[thick,fill=white] (.6,.1) circle (3pt);
        \node[font=\scriptsize] (P2) at (.5,-.2) {$[4k-3,3]^{\oplus N}$};
        \draw[thick,fill=white] (-.6,-.75) circle (3pt);
        \node[font=\scriptsize] (P3) at (-.1,-.8) {$\veryred{[2^{2k}]}$};
    \end{tikzpicture}
\end{center}
\vspace{3mm}

\subcaption{$\mathcal{S}_{\mathfrak{so}_{4k}}\langle S^2 \rangle \{\veryred{[2^{2k}]}, \veryred{[2^{2k}]},\, [4k-3,3]^{\oplus N}\}$}
\end{subfigure}
\begin{subfigure}{0.5\textwidth}
\begin{center}
    \begin{tikzpicture}
        \shade[inner color=white, outer color=blue!40, opacity = 0.4] (0,0) circle (1.4cm);
        \draw (0,0) circle (1.4cm);
        \draw (-1.4,0) arc (180:360:1.4 and 0.6);
        \draw[dashed] (1.4,0) arc (0:180:1.4 and 0.6);
        \draw[thick,fill=white] (-.6,.75) circle (3pt);
        \node[font=\scriptsize] (P1) at (-.1,.8) {$\veryred{[2^{2k}]}$};
        \draw[thick,fill=white] (0,.1) circle (3pt);
        \draw[thick,fill=black] (.2,.1) circle (.5pt);
        \draw[thick,fill=black] (.3,.1) circle (.5pt);
        \draw[thick,fill=black] (.4,.1) circle (.5pt);
        \draw[thick,fill=white] (.6,.1) circle (3pt);
        \draw[thick,fill=white] (.6,.1) circle (3pt);
        \node[font=\scriptsize] (P2) at (.5,-.2) {$[4k-3,3]^{\oplus N}$};
        \draw[thick,fill=white] (-.6,-.75) circle (3pt);
        \node[font=\scriptsize] (P3) at (-.05,-.8) {$\veryblue{[2^{2k}]}$};
    \end{tikzpicture}
\end{center}
\vspace{3mm}

\subcaption{$\mathcal{S}_{\mathfrak{so}_{4k}}\langle S^2 \rangle \{\veryred{[2^{2k}]}, \veryblue{[2^{2k}]}, [4k-3,3]^{\oplus N}\}$}
\end{subfigure}
\vspace{2mm}
\caption{The $(N+2)$-punctured spheres associated to the class $\mathcal{S}$ theories where the Higgs branch operator spectrum differs as described in equations \eqref{eqn:2to2kredred} and \eqref{eqn:2to2kredblue}. The former contains a $\hat{B}_{(k-1)(N-1)-1}$ operator that the latter does not.}
\label{fig:2to2k}
\end{figure}
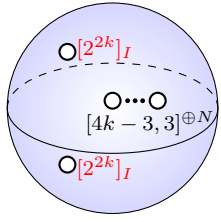
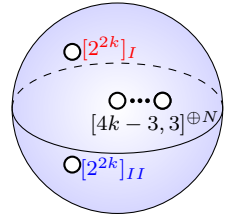

Since we take $N$ of the punctures to be simple, $[4k-3,3]$, equation \eqref{nLambdadef} reduces to
\begin{equation}\label{nLambdaspecialized}
    n_\Lambda= \Lambda\cdot C^{-1}\cdot\Bigl((0,0,\dots,0,2N;0,0)-w(O_1)-w(O_2)\Bigr) \,.
\end{equation}
In the examples that we wish to consider, we have
\begin{equation}\label{weightedexamples}
    \begin{split}
        w\bigl(\veryred{[2^{2k}]}\bigr)&=\bigl(0,0,\dots,0;(1+(-1)^k),(1-(-1)^k)\bigr) \,,\\
        w\bigl(\veryblue{[2^{2k}]}\bigr)&=\bigl(0,0,\dots,0;(1-(-1)^k),(1+(-1)^k)\bigr) \,,\\
        w\bigl(\veryred{[4^2,2^{2k-4}]}\bigr)&=\bigl(0,2,0,\dots,0;(1+(-1)^k),(1-(-1)^k)\bigr) \,,\\
        w\bigl(\veryblue{[4^2,2^{2k-4}]}\bigr)&=\bigl(0,2,0,\dots,0;(1-(-1)^k),(1+(-1)^k)\bigr) \,,\\
        w\bigl(\veryred{[(2k-2l)^2,2^{2l}]}\bigr)&=\bigl(\underbrace{0,2,0,2,\dots,0,2}_{2(k-l-1)},\underbrace{0,0,\dots,0}_{2l};(1+(-1)^k),(1-(-1)^k)\bigr) \,,\\
        w\bigl(\veryblue{[(2k-2l)^2,2^{2l}]}\bigr)&=\bigl(\underbrace{0,2,0,2,\dots,0,2}_{2(k-l-1)},\underbrace{0,0,\dots,0}_{2l};(1-(-1)^k),(1+(-1)^k)\bigr) \,.
    \end{split}
\end{equation}
We want to extract the values of $n_{S^+}$ and $n_{S^-}$, for which we need
\begin{equation}\label{SCinv}
    \begin{split}
        S^+\cdot C^{-1} &= \tfrac{1}{2}(1,2,3,\dots,2k-2;k,k-1) \,,\\
        S^-\cdot C^{-1} &= \tfrac{1}{2}(1,2,3,\dots,2k-2;k-1,k) \,.
    \end{split}
\end{equation}
Putting together equations \eqref{nLambdaspecialized}, \eqref{weightedexamples}, and \eqref{SCinv}, we get $n_{S^+}$ and $n_{S^-}$ for any pair of punctures $O_1$ and $O_2$. For instance, we see that 
\begin{equation}\label{eqn:2to2kredred}
    \begin{aligned}
        n_{S^+}\bigl(\veryred{[2^{2k}]},\veryred{[2^{2k}]}\bigr)&=2(k-1)(N-1)-\bigl(1+(-1)^k\bigr) \,, \\
        n_{S^-}\bigl(\veryred{[2^{2k}]},\veryred{[2^{2k}]}\bigr)&=2(k-1)(N-1)-\bigl(1-(-1)^k\bigr) \,,
    \end{aligned}
\end{equation}
whereas we get different values for the other theories:
\begin{equation}\label{eqn:2to2kredblue}
    \begin{aligned}
        n_{S^+}\bigl(\veryred{[2^{2k}]},\veryblue{[2^{2k}]}\bigr)&=2(k-1)(N-1)-1 \,, \\
        n_{S^-}\bigl(\veryred{[2^{2k}]},\veryblue{[2^{2k}]}\bigr)&=2(k-1)(N-1)-1 \,.
    \end{aligned}
\end{equation}
Thus we find that the discrepancy between the theories $\mathcal{S}_{\mathfrak{so}_{4k}}\langle S^2 \rangle \{\veryred{[2^{2k}]}, \veryred{[2^{2k}]},\, [4k-3,3]^{\oplus N}\}$ and \ifpreprintoption\else\goodbreak\fi $\mathcal{S}_{\mathfrak{so}_{4k}}\langle S^2 \rangle \{\veryred{[2^{2k}]}, \veryblue{[2^{2k}]}, [4k-3,3]^{\oplus N}\}$ first appears at order $\tau^{2(k-1)(N-1)-2}$ where the representation $\Lambda=S^+$ (for $k$ even) or $\Lambda=S^-$ (for $k$ odd) contributes a $\hat{B}_{(k-1)(N-1)-1}$ operator. We depict these two distinct class $\mathcal{S}$ theories in terms of their $N+2$ punctured spheres in Figure \ref{fig:2to2k}. 

In fact, from equation \eqref{Rlambdadef}, we see that this operator is a singlet of the flavor symmetry. This is precisely the state that was determined, in equation \eqref{eqn:eg1}, to exist in the nilpotent Higgsing of the 6d rank $N$ $(\mathfrak{so}_{4k}, \mathfrak{so}_{4k})$ conformal matter theory by the pair of nilpotent orbits $(\veryred{[2^{2k}]}, \veryred{[2^{2k}]})$, and which does not exist for the $(\veryred{[2^{2k}]}, \veryblue{[2^{2k}]})$ Higgsing. In this way, we see that the class $\mathcal{S}$ analysis of the Higgs branch confirms the conclusion of the 6d $(1,0)$ analysis.

Applying this method to all pairs of very even D-partitions that were studied in Section \ref{sec:examples}, we observe that the 4d class $\mathcal{S}$ and 6d $(1,0)$ approaches to the Higgs branch agree, as (perhaps) expected. The most general pair of very even D-partitions, studied at the end of Section \ref{eqn:lasteg} was
\begin{equation}
    O = [(2k-2\ell)^2, 2^{2\ell}] \,, \quad O'=[(2k-2\ell^\prime)^2, 2^{2\ell^\prime}] \,.
\end{equation}
Combining equations \eqref{nLambdaspecialized}, \eqref{weightedexamples}, and \eqref{SCinv} we can again determine $n_{S^\pm}$ for the pairs $(\veryred{O},\veryred{O'})$ and $(\veryred{O},\veryblue{O'})$
We find for the pairs of two reds $(\veryred{O},\veryred{O'})$,
\begin{equation}
    \begin{aligned}
        n_{S^+}\bigl(\veryred{O},\veryred{O'}\bigr)&= 2N(k-1) - (k - \ell)^2 - (k - \ell^\prime)^2 - \ell - \ell^\prime+\bigl(1-(-1)^k\bigr) \,, \\
        n_{S^-}\bigl(\veryred{O},\veryred{O'}\bigr)&=2N(k-1) - (k - \ell)^2 - (k - \ell^\prime)^2 - \ell - \ell^\prime+\bigl(1+(-1)^k\bigr) \,,
    \end{aligned}
\end{equation}
and for the pairs of a red and a blue $(\veryred{O},\veryblue{O'})$,
\begin{equation}
    \begin{aligned}
        n_{S^+}\bigl(\veryred{O},\veryblue{O'}\bigr)&= 2N(k-1) - (k - \ell)^2 - (k - \ell^\prime)^2 - \ell - \ell^\prime+1 \,, \\
        n_{S^-}\bigl(\veryred{O},\veryblue{O'}\bigr)&=2N(k-1) - (k - \ell)^2 - (k - \ell^\prime)^2 - \ell - \ell^\prime+1 \,.
    \end{aligned}
\end{equation}
The Hall--Littlewood indices of the two theories begin to differ at order 
\begin{equation}
    \tau^{2N(k-1) - (k - \ell)^2 - (k - \ell^\prime)^2 - \ell - \ell^\prime} \,,
\end{equation}
where there is one additional flavor singlet Higgs branch operator in the $(\veryred{\text{red}}, \veryred{\text{red}})$ theory that is absent in the $(\veryred{\text{red}}, \veryblue{\text{blue}})$ theory. This is identical with the result from the 6d $(1,0)$ tensor branch analysis as given in equation \eqref{eqn:boo}. Using the methodologies described throughout this paper, the extension to an arbitrary pair of very even D-partitions, both on the 6d $(1,0)$ and class $\mathcal{S}$ sides, is straightforward, though somewhat tedious.

\section{Conclusion}\label{sec:conc}

In this paper, we have demonstrated that distinct 6d $(1,0)$ SCFTs can share the same description of the low-energy theory that lives at the generic point of the tensor branch. Such SCFTs differ in their spectrum of Higgs branch operators, which we compute in two independent ways for the very even nilpotent Higgsing of rank $N$ $(D, D)$ conformal matter. First, we consider the generators of the Higgs branch spectrum of the building blocks, i.e.~minimal $(D, D)$ conformal matter, out of which the 6d SCFTs we consider are built; we then construct gauge-invariant operators out of these generators. Alternatively, we consider the compactification on a torus, which preserves the Higgs branch, and compute the Higgs branch spectrum from the dual class $\mathcal{S}$ description of the resulting 4d $\mathcal{N}=2$ SCFTs. Both approaches lead to identical results.

To conclude, we give several examples which demonstrate how the two possibilities for compositing via minimal $(D, D)$ conformal matter can lead to distinct 6d SCFTs with the same tensor branch description, outside of the class of theories obtained via very even nilpotent Higgsing of rank $N$ $(D, D)$ conformal matter. In these examples, there generally do not exist known class $\mathcal{S}$ duals, and thus the powerful techniques used in this paper to verify the computation of Higgs branch operators cannot be applied. However, we seek to emphasize that such compositing may also be ambiguous beyond minimal $(D, D)$ conformal matter, for example, in configurations where the compositing theory is instead $\overset{\mathfrak{su}_K}{1}$. Further analysis is required to determine when the tensor branches constructed via the algorithms of \cite{Heckman:2013pva,Heckman:2015bfa} correspond to multiple 6d $(1,0)$ SCFTs.

\subsection{Flavor algebras from E-strings}

Flavor symmetries in 6d $(1,0)$ SCFTs can arise in a variety of ways, as pointed out in, for example, \cite{Frey:2018vpw,Baume:2021qho}. One particular source is the so-called ``E-string flavor'', which occurs when we have a configuration of the form
\begin{equation}
    \cdots \overset{\mathfrak{g}}{m}\ 1_\rho\ \overset{\mathfrak{h}}{n} \cdots \,.
\end{equation}
We write a subscript $\rho$ on the $(-1)$-curve to stress that compositing the $(-m)$- and $(-n)$-curves together via an E-string involves a choice of embedding 
\begin{equation}
    \rho \, : \, \mathfrak{g} \oplus \mathfrak{h} \rightarrow \mathfrak{e}_8 \,.
\end{equation}
The non-Abelian part of the flavor algebra arising from such a compositing is 
\begin{equation}
    \mathfrak{f}^\text{E-string} = \operatorname{Commutant}(\rho, \mathfrak{g} \oplus \mathfrak{h}) \,,
\end{equation}
i.e.~the commuting subalgebra of the gauge symmetries under the embedding $\rho$, which is highly dependent on the choice of $\rho$. In Section \ref{sec:so8}, we explore the tensor branch configuration 
\begin{equation}
    \overset{\mathfrak{so}_7}{3}\ 1_{\rho}\ \overset{\mathfrak{so}_7}{3} \,,
\end{equation}
and we discover that there are two inequivalent $\rho$, which have commutants $\mathfrak{u}_1$ and $\varnothing$, respectively. The flavor symmetry is an invariant of the SCFT, thus these two tensor branch configurations correspond to distinct SCFTs, and the duality to class $\mathcal{S}$ verifies that both theories exist as interacting 6d $(1,0)$ SCFTs. Such ambiguity in the choice of embedding is ubiquitous in the geometric constructions of \cite{Heckman:2013pva,Heckman:2015bfa}, and raises the question of which embeddings lead to interacting SCFTs at the origin of the tensor branch. We explore an example tensor branch configuration with such an ambiguity that appears in \cite{Frey:2018vpw}. Consider
\begin{equation}\label{eqn:ambigtens}
    [\mathfrak{so}_7]\ \overset{\mathfrak{su}_2}{2}\ 1_\rho\ \overset{\mathfrak{so}_8}{4} \,.
\end{equation}
To understand how the $(-1)$-curve composites between the two neighboring curves, we need to understand embeddings
\begin{equation}\label{eqn:concembed}
    \rho \, : \, \mathfrak{su}_2 \oplus \mathfrak{so}_8 \oplus \mathfrak{f}^\text{E-string} \rightarrow \mathfrak{e}_8 \,,
\end{equation}
where the $\mathfrak{su}_2$ and $\mathfrak{so}_8$ factors must have Dynkin embedding index $1$ as they are gauged. For a detailed study of the relevance of Dynkin embedding index one for F-theory compactifications see \cite{Esole:2020tby}. In \cite{Frey:2018vpw},  two embeddings were pointed out, with
\begin{equation}
  \mathfrak{f}^\text{E-string} = \mathfrak{su}_2^{\oplus 3} \qquad \text{ and } \qquad \mathfrak{f}^\text{E-string} = \mathfrak{sp}_2 \,,
\end{equation}
however, in principle, there may be additional embeddings. In fact, the embedding 
\begin{equation}
    \rho \,: \, \mathfrak{su}_2 \oplus \mathfrak{so}_8 \oplus \mathfrak{sp}_2 \rightarrow \mathfrak{e}_8 \,,
\end{equation}
is not appropriate in this configuration, as the Dynkin embedding index of the $\mathfrak{su}_2$ factor is $2$, not $1$. This tensor branch configuration can also be obtained from nilpotent Higgsing of rank two $(\mathfrak{e}_6, \mathfrak{e}_6)$ conformal matter:
\begin{equation}
    \underset{[\mathfrak{e}_6]} {1}\overset{\mathfrak{su}_3}{3}\ 1\ \overset{\mathfrak{e}_6}{6}\ 1\ \overset{\mathfrak{su}_3}{3}\underset{[\mathfrak{e}_6]}{1} \quad \xrightarrow{\quad(2A_1, D_4(a_1))\quad} \quad \underset{[\mathfrak{so}_7]} {\overset{\mathfrak{su}_2}{2}}1\ \overset{\mathfrak{so}_8}{4} \,.
\end{equation}
Compactification of at least one of the SCFTs associated to the tensor branch configuration in equation \eqref{eqn:ambigtens} then has a dual description in terms of class $\mathcal{S}$ of type $\mathfrak{e}_6$ on a sphere with two simple punctures and two punctures associated to the nilpotent orbits $2A_1$ and $D_4(a_1)$.\footnote{For the exceptional Lie algebras we use the Bala--Carter notation \cite{MR417306,MR417307} for the nilpotent orbits, see \cite{MR1251060} for more details.} From the class $\mathcal{S}$ perspective, the Hall--Littlewood index yields
\begin{equation}
    1 + 30\tau^2 +64 \tau^3 + \mathcal{O}(\tau^4) \,,
\end{equation}
which demonstrates that the flavor algebra is enhanced from the manifest $(\mathfrak{so}_7)_{16} \oplus \mathfrak{u}_1^{\oplus 3}$ to\footnote{The levels of the enhanced $\mathfrak{su}_2$ factors are not immediately obvious. They can be obtained by considering the degeneration limit of the 4-punctured sphere in which the theory becomes a weakly-coupled $SU(2)$ gauging of interacting fixture \#43 of \cite{Chacaltana:2014jba} with an additional half-hypermultiplet in the $\bm{2}$. Alternatively, the S-dual realization is a $Spin(8)$ gauging of the $(E_7)_{24}\times Spin(7)_{16}$ SCFT, which is interacting fixture \#3 of \cite{Chacaltana:2014jba}. The centralizer of $\mathfrak{so}_8\subset (\mathfrak{e}_7)_{24}$ is $(\mathfrak{su}_2)_{24}^{\oplus3}$.}
\begin{equation}
    \mathfrak{f} = (\mathfrak{so}_7)_{16} \oplus (\mathfrak{su}_2)_{24}^{\oplus 3} \,.
\end{equation}
Thus, we can confidently state, via the duality to class $\mathcal{S}$, that 
\begin{equation}
    \underset{\vphantom{{T}_T^T}[\mathfrak{so}_7]}{ \overset{\mathfrak{su}_2}{2}}\;\underset{[\mathfrak{su}_2^{\oplus 3}]}{1}\overset{\mathfrak{so}_8}{4} \,,
\end{equation}
describes the tensor branch of an interacting 6d $(1,0)$ SCFT, however, this does \emph{not} suggest that any other embeddings of the form in equation \eqref{eqn:concembed} do not give rise to interacting 6d SCFTs. It is an important question for the understanding of the landscape of possible 6d $(1,0)$ SCFTs to determine if tensor branch geometries like that in equation \eqref{eqn:ambigtens} correspond to one or more interacting SCFTs.

\subsection{From nilpotent orbits to \texorpdfstring{$E_8$}{E₈}-homomorphisms}

The rank $N$ $(\mathfrak{g}, \mathfrak{g})$ conformal matter theories have many tools with which their properties can be studied. In particular, they can be realized as the worldvolume theories on a stack of M5-branes probing a $\mathbb{C}^2/\Gamma_{\mathfrak{g}}$ orbifold. Here, $\Gamma_\mathfrak{g}$ is the finite subgroup of $SU(2)$ of the same ADE-type as $\mathfrak{g}$. In this M-theory framework, the 6d SCFT behaves as a defect in 7d super Yang--Mills, with gauge algebra $\mathfrak{g}$, and the SCFT thus inherits a $\mathfrak{g} \oplus \mathfrak{g}$ flavor symmetry. Each flavor symmetry factor can be Higgsed by a nilpotent orbit of $\mathfrak{g}$, corresponding to turning on asymptotic boundary conditions for the scalar inside of the 7d vector multiplet. Nilpotent orbits of simple Lie algebras are well-studied and classified \cite{MR417306,MR417307}. In addition, when compactified on a torus the resulting 4d $\mathcal{N}=2$ SCFTs have an alternative description in terms of class $\mathcal{S}$ and thus one has an independent construction with which to study aspects of the Higgs branch of the 6d $(1,0)$ SCFTs.

For other classes of 6d $(1,0)$ SCFTs, we are not so lucky. In this section, we remark on the 6d SCFTs obtained via Higgs branch renormalization group flows from the rank $N$ $(\mathfrak{e}_8, \mathfrak{g})$ orbi-instanton theories. The orbi-instanton is realized in M-theory as a stack of $N$ M5-branes probing a $\mathbb{C}^2/\Gamma_\mathfrak{g}$ orbifold singularity, and contained inside of an M9-brane \cite{DelZotto:2014hpa}.  In such a configuration, one must specify the boundary conditions inside of the M9-brane, which are fixed by a choice of homomorphism $\pi:\, \Gamma_\mathfrak{g} \rightarrow E_8$, and changing these boundary conditions corresponds to performing Higgs branch renormalization group flows. We can consider the Higgsed rank $N$ $(\mathfrak{e}_8, \mathfrak{g})$ orbi-instanton theories as 
\begin{equation}
    \Omega_{\mathfrak{g},N}(\pi, \sigma) \,,
\end{equation}
where $\pi :\Gamma_\mathfrak{g} \rightarrow E_8$ is the $E_8$-homomorphism with which the $\mathfrak{e}_8$ flavor symmetry is Higgsed, and $\sigma :\, \mathfrak{su}_2 \rightarrow \mathfrak{g}$ is the nilpotent orbit by which the $\mathfrak{g}$ flavor symmetry is Higgsed. Homomorphisms from $\Gamma_{\mathfrak{su}_K}$ and $\Gamma_{\mathfrak{e}_8}$ to $E_8$ have been classified in \cite{MR739850} and \cite{MR1839999}, respectively, however in each of the other cases there is no known complete classification. In addition, the orbi-instantons and their Higgsings do not generally have known class $\mathcal{S}$ descriptions after compactification on a torus, and thus that avenue for understanding the 6d Higgs branch is closed. 

In \cite{Frey:2018vpw}, the authors seek to classify the homomorphims $\Gamma_\mathfrak{g} \rightarrow E_8$ utilizing the study of 6d $(1,0)$ tensor branch geometries. Of course, if one were to attempt to derive the nilpotent orbits in the same way, and one did not know about the two different ways of compositing using $\overset{\mathfrak{sp}_q}{1}$ pointed out in this paper, then one would not notice that each very even D-partition corresponds to \emph{two} nilpotent orbits! This is pointed out in \cite{Frey:2018vpw}, where the authors claim only to classify such homomorphims only up to outer automorphism. Here, we argue that by including the information about different possible compositings in the tensor branch description, one can see the difference between $E_8$-homomorphisms that appear to correspond to the same tensor branch. Furthermore, one should again be able to determine properties of the Higgs branches of the two SCFTs obtained in such a way, and observe how they differ.

For conciseness, we consider the following example: take the rank $N$ $(\mathfrak{e}_8, \mathfrak{so}_8)$ orbi-instanton, which has the following tensor branch configuration:
\begin{equation}
    [\mathfrak{e}_8]\,1\ 2\ \overset{\mathfrak{su}_2}{2}\ \overset{\mathfrak{g}_2}{3}\ \overbrace{1\overset{\mathfrak{so}_8}{4}\ 1\cdots \overset{\mathfrak{so}_8}{4}\ 1}^{N-1\,\,  (-4)\text{-curves}}\,[\mathfrak{so}_8] \,.
\end{equation}
To be more illuminating, we focus on the case where $N = 3$. Consider the homomorphism $\pi:\, \Gamma_{\mathfrak{so}_8} \rightarrow E_8$, discussed in \cite{Frey:2018vpw}, which triggers a Higgs branch renormalization group flow to a new 6d SCFT with the following tensor branch description:
\begin{equation}
    [\mathfrak{e}_8]\,1\ 2\ \overset{\mathfrak{su}_2}{2}\ \overset{\mathfrak{g}_2}{3}\ 1\ \overset{\mathfrak{so}_8}{4}\ 1\ \overset{\mathfrak{so}_8}{4}\ 1\,[\mathfrak{so}_8] \quad\overset{\pi}{\longrightarrow}\quad [\mathfrak{so}_9]\,1\ \underset{[\mathfrak{sp}_2]}{\overset{\mathfrak{so}_7}{3}}\ 1\ \overset{\mathfrak{so}_8}{4}\ 1\,[\mathfrak{so}_8] \,.
\end{equation}
As we can see, there is a $(-1)$-curve, corresponding to the E-string, that composites between the $\overset{\mathfrak{so}_7}{3}$ and $\overset{\mathfrak{so}_8}{4}$ building blocks. In Section \ref{sec:so8}, we saw that such a compositing was ambiguous, and could lead to distinct 6d SCFTs. Due to this ambiguity, we propose that there are two distinct $E_8$-homomorphisms, labeled as $\veryred{\pi}$ and $\veryblue{\pi}$, which lead to the same tensor branch. Further Higgsing the $\mathfrak{so}_8$ flavor symmetry on the right by either the $\veryred{[2^4]}$ or the $\veryblue{[2^4]}$ nilpotent orbit leads to the following tensor branch configuration:
\begin{equation}
 [\mathfrak{so}_9]\,1\ \underset{[\mathfrak{sp}_2]}{\overset{\mathfrak{so}_7}{3}}\ 1\ \overset{\mathfrak{so}_8}{4}\ 1\,[\mathfrak{so}_8] \quad\rightarrow\quad [\mathfrak{so}_9]\,1\ \underset{[\mathfrak{sp}_2]}{\overset{\mathfrak{so}_7}{3}}\ 1\ \overset{\mathfrak{so}_7}{3}\,[\mathfrak{sp}_2] \,.
\end{equation}
Based on the arguments in Section \ref{sec:so8}, in particular the two distinct embeddings of $\mathfrak{so}_7 \oplus \mathfrak{so}_7$ inside of $\mathfrak{e}_8$ given in equation \eqref{eqn:so7dec}, we expect that this tensor branch configuration corresponds to \emph{two} distinct 6d SCFTs; $(\veryred{\pi}, \veryred{[2^4]})$ with an additional $\mathfrak{u}_1$ flavor symmetry, and $(\veryblue{\pi}, \veryred{[2^4]})$ without. Unfortunately, as we discussed, there does not exist known class $\mathcal{S}$ duals for the compactifications of Higgsed orbi-instantons on $T^2$, and thus, unlike in the study of Higgsed conformal matter, we lack an independent method to verify this plurality of 6d $(1,0)$ SCFTs.\footnote{When $\mathfrak{g} = \mathfrak{su}_K$ there are class $\mathcal{S}$ descriptions for the Higgsed orbi-instantons \cite{Mekareeya:2017jgc}. In that reference, the authors point out that there exist $E_8$-homomorphisms such that after Higgsing the tensor branch has the form $\overset{\mathfrak{sp}_q}{1}\,\overset{{su}_{2q+8}}{2}\cdots$, and it is argued that, in such cases, the choice of $\theta$-angle for the $\mathfrak{sp}_q$ gauge algebra can lead to two distinct 6d SCFTs. It would be interesting to verify this from the class $\mathcal{S}$ perspective.}

\subsection*{Acknowledgements}

We thank Jonathan Heckman for comments on an earlier draft of this manuscript. 
The authors thank the  Aspen Center for Physics, which is supported by National Science Foundation grant PHY--1607611, for hospitality during the intermediate stage of this work. C.L.~also thanks the California Institute of Technology for hospitality during the initial stage of this work. The work of J.D.~is supported in part by the National Science Foundation under Grant No.~PHY--1914679. M.J.K.~is partially supported by a Sherman Fairchild Postdoctoral Fellowship and the U.S.~Department of Energy, Office of Science, Office of High Energy Physics, under Award Number DE-SC0011632. C.L.~acknowledges support from DESY (Hamburg, Germany), a member of the Helmholtz Association HGF. 

\bibliography{references}{}
\bibliographystyle{sortedbutpretty}
\end{document}